\documentclass[a4paper,fleqn,usenatbib]{mnras}

\usepackage{acronym}
\usepackage[usenames,dvipsnames]{xcolor}
\usepackage{amsfonts,amsmath,amssymb}
\usepackage{commath}
\usepackage{booktabs}
\usepackage{multirow}
\usepackage{graphicx}

\renewcommand*{\appref}[1]{\hyperref[#1]{Appendix~\ref{#1}}}
\hypersetup{
    linktocpage,
    colorlinks,
    linkcolor={blue!75!black},
    citecolor={blue!75!black},
    urlcolor={blue!75!black}
}

\DeclareMathOperator{\erf}{erf}
\newcommand*{\bigO}{\ensuremath{O}}
\newcommand*{\half}{\ensuremath{\tfrac{1}{2}}}
\newcommand*{\E}[2][]{\ensuremath{\operatorname{E}[{#2}]_{#1}}}
\newcommand*{\Var}[2][]{\ensuremath{\operatorname{Var}[{#2}]_{#1}}}
\newcommand*{\vectheta}{\ensuremath{\vec{\theta}}}
\newcommand*{\vecthetasq}{\ensuremath{\|\vectheta\|^2}}
\newcommand*{\actacc}{\ensuremath{\tau_\mathrm{acc}}}
\newcommand*{\actgeo}{\ensuremath{\tau_\mathrm{geo}}}
\newcommand*{\acttest}{\ensuremath{\tau_\mathrm{test}}}
\newcommand*{\logL}{\ensuremath{\log L}}
\newcommand*{\unit}[1]{\ensuremath{\,#1}}

\newcommand*{\Msun}{\unit{M}_{\odot}}
\newcommand*{\imrpp}{\mbox{\em IMRPhenomP}}
\newcommand*{\tft}{\mbox{\em TaylorF2}}
\newcommand*{\sttf}{\mbox{\em SpinTaylorT4}}
\newcommand*{\Mc}{\ensuremath{\mathcal{M}}}

\newcommand*{\emcee}{\textsc{emcee}}
\newcommand*{\lalinf}{\textsc{LALInference}}
\newcommand*{\emceeref}{\url{https://github.com/willvousden/ptemcee}}

\acrodef{MCMC}{Markov chain Monte Carlo}
\acrodef{RWMH}{random walk Metropolis Hastings}
\acrodef{ACT}{autocorrelation time}
\acrodef{KL}{Kullback--Leibler}
\acrodef{CBC}{compact binary coalescence}
\acrodef{GW}{gravitational wave}
\acrodef{LSC}{LIGO Scientific Collaboration}
\acrodef{SNR}{signal-to-noise ratio}
\acrodef{BNS}{binary neutron star}
\acrodef{BBH}{binary black hole}
\acrodef{PSD}{power spectral density}

\makeatletter
\def\maxwidth#1{\ifdim\Gin@nat@width>#1 #1\else\Gin@nat@width\fi}
\makeatother

\begin{document}

\title[Temperature dynamics for PTMCMC samplers]{Dynamic temperature selection
for parallel-tempering in Markov chain Monte Carlo simulations}

\author[W. D. Vousden, W. M. Farr and I. Mandel]
{W. D. Vousden\thanks{E-mail:
        will@star.sr.bham.ac.uk (WDV);
        w.farr@bham.ac.uk (WMF);
        imandel@star.sr.bham.ac.uk (IM)},
W. M. Farr\footnotemark[1] and
I. Mandel\footnotemark[1]\\
University of Birmingham, Edgbaston, Birmingham, B15 2TT, United Kingdom}

\date{\today}

\pagerange{\pageref{firstpage}--\pageref{lastpage}} \pubyear{2002}

\maketitle

\label{firstpage}

% vim: set ft=tex :

\begin{abstract}
    Modern problems in astronomical Bayesian inference require efficient methods
    for sampling from complex, high-dimensional, often multi-modal probability
    distributions.  Most popular methods, such as \acl*{MCMC} sampling, perform
    poorly on strongly multi-modal probability distributions, rarely jumping
    between modes or settling on just one mode without finding others.  Parallel
    tempering addresses this problem by sampling simultaneously with separate
    Markov chains from tempered versions of the target distribution with reduced
    contrast levels.  Gaps between modes can be traversed at higher
    temperatures, while individual modes can be efficiently explored at lower
    temperatures.  In this paper, we investigate how one might choose the ladder
    of temperatures to achieve more efficient sampling, as measured by the
    autocorrelation time of the sampler.  In particular, we present a simple,
    easily-implemented algorithm for dynamically adapting the temperature
    configuration of a sampler while sampling.  This algorithm dynamically
    adjusts the temperature spacing to achieve a uniform rate of exchanges
    between chains at neighbouring temperatures. We compare the algorithm to
    conventional geometric temperature configurations on a number of test
    distributions and on an astrophysical inference problem, reporting
    efficiency gains by a factor of 1.2--2.5 over a well-chosen geometric
    temperature configuration and by a factor of 1.5--5 over a poorly chosen
    configuration. On all of these problems a sampler using the dynamical
    adaptations to achieve uniform acceptance ratios between neighbouring chains
    outperforms one that does not.
\end{abstract}

% TODO: keywords?
\nokeywords

% vim: set ft=tex :

\section{Introduction}
\label{sec:intro}

Many problems in astronomical data analysis and Bayesian statistical inference
demand the characterisation of high-dimensional probability distributions with
complicated structures.  Lacking analytic forms, these distributions must be
explored numerically, usually via Monte Carlo methods.

Parallel tempered \ac{MCMC}, a development on standard \ac{MCMC}, uses several
Markov chains in parallel to explore a target distribution at different
``temperatures'' \citep{Earl2005,Swendsen1986,Geyer1991}.  As the temperature
increases, the posterior distribution asymptotes to the prior, allowing a chain
to efficiently explore the whole prior volume without becoming stuck in regions
of the parameter space with high probability density.  At lower temperatures, a
chain can more efficiently sample from such a high-probability region.
Meanwhile, exchange of positions between chains allows colder chains to migrate
between widely separated modes in the parameter space \citep{Geyer1991}.
Parallel tempered \ac{MCMC} samplers are thus particularly well-suited to
sampling posterior distributions with well-separated modes, where a regular
\ac{MCMC} sampler would take many iterations to find its way between modes.

An open problem in the application of parallel tempering is selecting a
specification, or ladder, of temperatures that minimises the \ac{ACT} of the
chain sampling the posterior distribution of interest.  The efficiency of a
given ladder hinges critically on the rate at which it can transfer the
positions in parameter space of samples between high and low temperatures.

In this paper we present a simple algorithm that adapts the temperature ladder
of an ensemble-based parallel tempered \ac{MCMC} sampler \citep{Goodman2010}
such that the rate of exchange between chains is uniform over the entire ladder.
The algorithm is easy to implement in existing code, and we provide an example
implementation for the \emcee{} sampler of \citet{emcee}.  We also present an
implementation for traditional, non-ensemble \ac{MCMC} samplers where a single
walker explores the parameter space.  We report favourable results from such an
implementation, along with a number of caveats.

In \autoref{sec:pt} we describe the parallel tempering formalism and lay out the
requirements for a good temperature ladder.  We discuss previous work on
temperature selection and suggest a definition of ladder optimality that, for
simple cases, proposes a geometric spacing of temperatures.  For illustration,
we apply these ideas in \autoref{sec:pt-ideal-gaussian} to the simple example of
an unbounded Gaussian posterior distribution.

In \autoref{sec:adaptive} we describe the algorithm mentioned above and then
apply it in \autoref{sec:tests} to a variety of test distributions.  We show
that, while our temperature selection strategy is not necessarily optimal in the
\ac{ACT} of the sampler, it nonetheless improves the \ac{ACT} compared to the
simple geometric spacing that is conventional in the literature \citep{Earl2005,
Sugita1999, Kofke2002, Kofke2004a} by factors of $>1.2$ for our test cases.

In \autoref{sec:gw} we apply our method to the astrophysically motivated -- and
more challenging -- problem of parameter estimation in the setting of \ac{GW}
data analysis, using a single-walker \ac{MCMC} sampler.  We demonstrate a
reduction in \ac{ACT} by as much as a factor of $2$ over a geometric ladder,
despite the caveats mentioned in \autoref{sec:discussion-caveats}.

We conclude in \autoref{sec:discussion} with a
discussion of our results and suggestions for further research.

\section{Parallel tempering}
\label{sec:pt}

Parallel tempering \citep{Earl2005,Swendsen1986,Geyer1991} is a development on
the standard \ac{MCMC} formalism that uses several Markov chains in parallel to
sample from tempered versions of the posterior distribution $\pi$,
\begin{equation}\label{eq:pt-posterior}
    \pi_T(\vectheta) \propto L(\vectheta)^{1/T}p(\vectheta)
    \text{,}
\end{equation}
where $L$ and $p$ are respectively the likelihood and prior distributions.

\newcommand\Tmin{\ensuremath{T_\mathrm{min}}}
\newcommand\Tmax{\ensuremath{T_\mathrm{max}}}
\newcommand\Tprior{\ensuremath{T_\mathrm{prior}}}

For high $T$, individual peaks in $L$ become flatter and broader, making the
distribution easier to sample via \ac{MCMC}.  A set of $N$ chains is assigned
temperatures in a ladder $T_1<T_2<\ldots<T_N$, with $T_1 = 1$ (the target
temperature).  The temperatures are typically geometrically spaced from 1 up to
some $\Tmax$, decided in advance (a convention that we shall discuss in more
detail in \autoref{sec:pt-ideal-gaussian}).

Each chain is allowed to explore its tempered distribution $\pi_T$ under an
\ac{MCMC} algorithm, while at pre-determined intervals ``swaps'' are proposed
between (usually adjacent\footnote{
    In principle, swaps can be proposed between any pair of chains.  However,
    since the swap acceptance ratio \eqref{eq:swap} decays exponentially with
    the separation of inverse temperatures, $\Delta\beta$, it is generally
    sufficient only to propose swaps between adjacent chains.
}) pairs of chains and accepted with probability
\begin{equation}\label{eq:swap}
    A_{i,j} = \min\left\{
                  \left(\frac{L(\vectheta_i)}{L(\vectheta_j)}\right)^{\beta_j-\beta_i},
                  \;
                  1
              \right\}
    \text{,}
\end{equation}
where $\vectheta_i$ is the current position in the parameter space of the
$i^\mathrm{th}$ chain and $\beta_i \equiv 1/T_i$ is the inverse temperature of
this chain.  When a swap is accepted, the chains exchange their positions in the
parameter space, so that chain $i$ is at $\vectheta_j$ and chain $j$ is at
$\vectheta_i$.  Since the hottest chains can access all of the modes of $\pi$
(as long as $\Tmax$ is chosen appropriately), their locations propagate to
colder chains, ultimately allowing the $T = 1$ (cold) chain to efficiently
explore the entire target distribution.  At the same time, the positions of the
colder chains propagate upward to higher temperature chains, where they are free
to explore the entire prior volume.

The goal in choosing an effective ladder of temperatures is to minimise the
\ac{ACT} of the cold chain (our measure of the efficiency of the sampler).  The
requirements to this end are two-fold:
\begin{enumerate}
    \item\label{item:condition-tmax} $\Tmax$ must be large enough that isolated
        modes of $L$ broaden sufficiently that an individual \ac{MCMC} chain can
        efficiently access all of these modes when sampling under the tempered
        posterior $\pi_T$ in \eqref{eq:pt-posterior} at $T = \Tmax$.  We denote
        this temperature $\Tprior$.
    \item\label{item:condition-spacing} Since $A_{i,j}$ depends on
        $\beta_i-\beta_j$, the differences between temperatures must be small
        enough that neighbouring chains can communicate their positions
        efficiently with one another.
\end{enumerate}
Both requirements depend sensitively on the (unknown) shape of the target
distribution, so it is difficult to select temperatures appropriately in
advance.

\begin{figure}
    \begin{center}
        \includegraphics[width=\maxwidth{\columnwidth}]{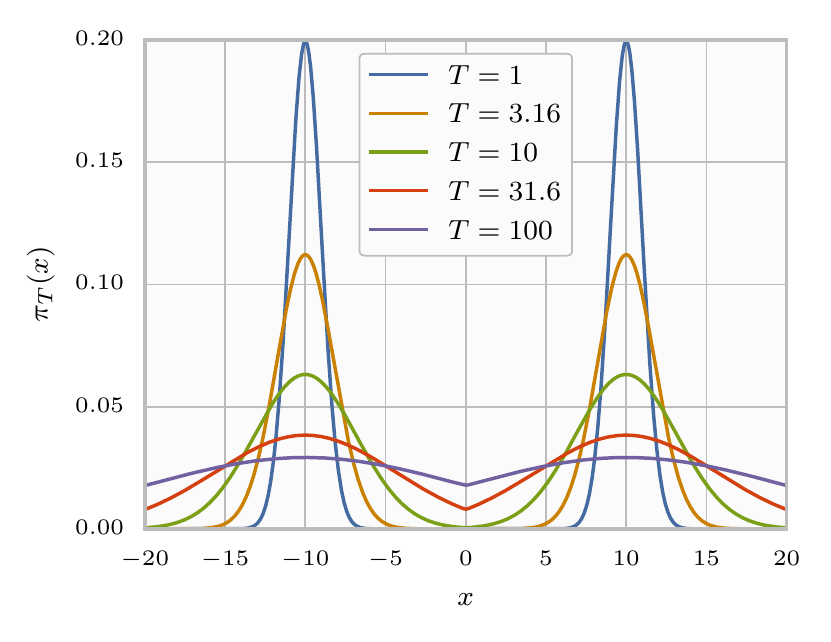}
        \caption[Tempered Gaussian distributions in one dimension]
        {\label{fig:tempered-gaussian}
            A one-dimensional target distribution with two Gaussian peaks of
            width $\sigma=1$ at $\mu=\pm 10$ normalised for a uniform prior over
            $[-20,20]$.  At $T=100$, the peaks broaden to $\sigma=10$, allowing
            an \ac{MCMC} chain sampling at this temperature to find both modes
            quickly, starting from anywhere within the prior volume.
        }
    \end{center}
\end{figure}

In choosing $\Tmax$, one must know roughly the relative size and separation of
the modes to be explored. As an example, consider a one-dimensional likelihood
with two Gaussian modes of width $\sigma=1$ and centres $\mu=\pm10$.  In order
to prevent a sampler from getting stuck on one of the modes, they must be
widened to roughly the separation between them\footnote{
    Ideally, the modes must also be widened enough that they extend to the edges
    of the prior volume.  A likelihood distribution with a single mode that
    occupies only a small fraction of the prior volume will take a long time to
    burn in.
}, giving $\sigma=\bigO(10)$.  The width of a Gaussian peak scales with the
temperature as $\sqrt{T}$, so we might choose $\Tmax=100$;
\autoref{fig:tempered-gaussian} illustrates the resulting coalescence of the
modes.  A different configuration of modes will, of course, require a different
$\Tmax$.

On the other hand, the swap acceptance probability $A_{i,j}$ depends on the
distribution of likelihood values at temperatures $T_i$ and $T_j$.  In the case
of a likelihood distribution comprising a single Gaussian mode, the
time-averaged acceptance ratios between chains, $\E{A_{i,j}}$, can be computed
analytically (see \autoref{sec:pt-ideal-gaussian}).

In general, we don't know in advance what the target distribution looks like,
and so choosing an effective ladder becomes a heuristic exercise, relying
largely on educated guesswork.  We are therefore motivated to find some method
of empirically determining an effective ladder.

\subsection{Ladder selection}
\label{sec:pt-ladder-selection}

For an $n$-dimensional problem, the conventional choice of temperatures is a
geometrically spaced ladder constructed so that approximately $23\%$ of swaps
proposed between chains will be accepted when sampling from an $n$-dimensional,
unbounded Gaussian distribution \citep{Earl2005, Braak2008, Roberts1998}.  We
shall discuss this convention in more detail in \autoref{sec:pt-ideal-gaussian}.

A consequence of this strategy is that increasing the number of chains $N$ does
not improve communication between existing chains, which is determined by
$\E{A_{i,j}} = 0.23$.  Instead, adding new chains extends the ladder to higher
temperatures.  This may be appropriate for an unbounded posterior, but for a
realistic problem with a finite prior volume, the acceptance ratio between
adjacent chains saturates to $\sim 100\%$ at some temperature $\Tprior$, at
which the posterior $\pi_T$ begins to look like the prior $p$.

\newcommand{\Nopt}{\ensuremath{{N_\mathrm{opt}}}}
For this geometric spacing scheme -- where $\Tprior$ is unknown -- there is
therefore an optimal number of chains, $\Nopt$, such that $\Tprior \approx
T_\Nopt \equiv \Tmax$.  For $N < \Nopt$ none of the chains will be sampling from
the prior (so the sampler may not find all of the modes), while for $N > \Nopt$
we end up with several chains sampling redundantly from the prior.

Since we are generally ignorant of $\Tprior$ for the problem at hand, we are
motivated to find an alternative temperature selection strategy.

It has been suggested in the literature \citep{Earl2005, Sugita1999, Kofke2002,
Kofke2004a} that one could select temperatures such that the acceptance ratios
$A_{i,j}$ are uniform for all pairs $(i,j)$ of adjacent chains, in an attempt to
ensure that each sample sequence $\vectheta(t)$ for $t = 1,2,\ldots$, as it
moves between chains, spends an equal amount of time at every temperature.
\citet{Sugita1999} justify this notion experimentally -- in the context of
molecular dynamics -- with test cases in which such a ladder indeed performs
well.  They use an algorithm derived from that of \citet{Hukushima1996}, which
selects temperatures according to an iterative process for which a uniform-$A$
ladder is a fixed point.  \citet{Earl2005} provide further references for
similar methods of determining temperature ladders that yield a given a target
acceptance ratio \citep{Rathore2005,Sanbonmatsu2002,Schug2004}.  However, these
methods do not address requirement~\ref{item:condition-tmax}, discussed above,
that the temperature ladder should reach a $\Tmax$ sufficient for all of the
modes of $L$ to mix (specified by $\Tprior$).

\citet{Kofke2002} discusses the selection of temperature ladders in the context
of molecular simulations.  He shows that, in simulations of such thermodynamic
systems, there is a close relation between the specific heat of the system,
$C_V$, and the acceptance ratios between adjacent temperatures.  In particular,
when $C_V$ is constant with respect to $T$ over a given temperature interval,
then a geometric spacing of temperatures on that interval yields uniform
acceptance ratios between adjacent temperatures.

In the language of thermodynamics, the energy of the system, $U$, is analogous
to $-\logL$, and an analogue to the specific heat can therefore be defined as
\begin{equation}\label{eq:specific-heat}
    C_V(T) = -\od{}{T} \E[T]{\logL}
    \text{,}
\end{equation}
where $\E[T]{\,\cdot\,}$ denotes the expectation operator over $\vectheta$ under
the distribution $\pi_T(\vectheta)$.  $\E[T]{\logL}$ is therefore the
expectation of the {\em untempered} log likelihood collected when sampling from
the posterior at temperature $T$.

In the context of Bayesian inference, \citeauthor{Kofke2002}'s result therefore
tells us that if the mean log likelihood collected by a sampler responds
linearly to changes in temperature, then a geometrically spaced temperature
ladder will achieve uniform acceptance ratios between adjacent chains.
Conversely, temperature intervals on which $\E[T]{\logL}$ is strongly non-linear
in $T$ represent a phase transition that will require more careful placement of
temperatures, as we shall show in \autoref{sec:tests}.

\subsection{The ideal Gaussian distribution: a simple example}
\label{sec:pt-ideal-gaussian}

In the simple case of a unimodal Gaussian likelihood under a flat prior, the
optimal temperature spacing at low temperatures -- where very little likelihood
mass is truncated by the prior -- can be analysed by approximating the prior to
be unbounded\footnote{
    The approximation breaks down at higher temperatures, where boundary effects
    become significant.  Indeed, with no prior boundaries, there is no $\Tprior$
    at which the mode is spread over the entire prior volume.
}.
We show that, for this tractable example, a geometric temperature spacing is
consistent with both the uniform-$A$ criterion and also with the alternative
criterion that the \ac{KL} divergence is uniform between all pairs of adjacent
chains.  We use the example to illustrate the relationship between the
analytical distribution of $\logL$, the acceptance ratio $A_{i,j}$, and the
temperature $T$.

\newcommand{\halfn}{\ensuremath{{\frac{n}{2}}}}
\newcommand{\ptilde}{\ensuremath{\tilde{p}}}
We shall work with an $n$-dimensional unit Gaussian centred on the origin (the
same result can be achieved for a general Gaussian through a simple change of
coordinates).  Since the prior is uniform and unbounded, we can restrict
attention to the likelihood distribution $L$.  In this case, the probability
density $\ptilde$ for the values of $\logL(\vectheta)$ collected by the sampler
is
\begin{equation}\label{eq:n-gauss-log-pi}
    \ptilde(\logL) = \frac{e^{\logL}(-\logL)^{\halfn-1}}
                         {\Gamma(\halfn)}
    \text{,}
\end{equation}
where $L$ is normalised so that $\logL(\vec{0}) = 0$ and $n$ is the
number of parameters.

At a temperature $T$, $-\logL$ simply follows a gamma distribution
$\Gamma(\alpha,\beta)$ with shape parameter $\alpha = n/2$ and rate parameter
$\beta = 1/T$. Thus, for a chain sampling at temperature $T$, the log likelihood
distribution is $\ptilde_T(\logL) = T \, \ptilde(\logL / T)$.

\begin{figure}
    \begin{center}
        \includegraphics[width=\maxwidth{\columnwidth}]{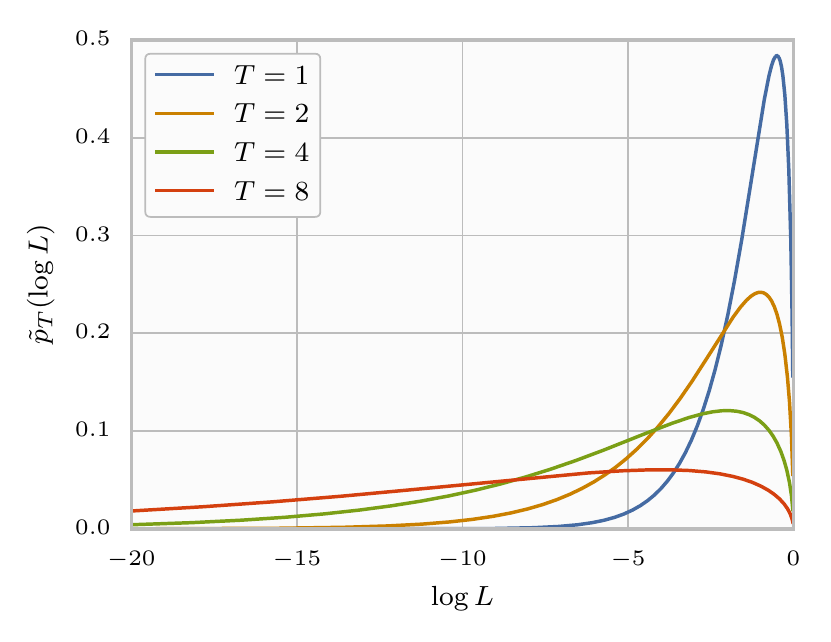}
        \caption[Tempered log-likelihood distributions]
        {\label{fig:ideal-gaussian-logl}
            The distribution of $\logL$ under a three-dimensional, unimodal
            Gaussian at various temperatures, where $L$ is normalised so that
            $\logL(\vec{0}) = 0$.  As $T \to \infty$, the variance of $\logL$
            diverges.  The legend is ordered to match the vertical order of the
            lines' peaks.
        }
    \end{center}
\end{figure}

\newcommand{\hgeo}{\ensuremath{{}_2\tilde{F}_1}}
Over long time-scales, the average acceptance ratio between chains $i$ and $j$
is
\begin{equation}\label{eq:acceptance-solution}
    \begin{split}
        \E{A_{i,j}} & = \iint\limits_{(-\infty,0]^2}
                             A_{i,j} \,
                             \ptilde_{T_i}(\logL_i) \, \ptilde_{T_j}(\logL_j) \
                             \dif\logL_i \, \dif\logL_j
                    \\
                    % Now for the big, long expression:
                    & =
                    \left(\frac{1}{\sqrt{\pi}} \, 2^{n-1} \gamma_{i,j} ^{-n/2} \, \Gamma \left(\frac{n+1}{2}\right)\right)
                    \\
                    & \mathrel{\phantom{=}}
                        \mathrel{\cdot} \Biggl(
                            \hgeo\left(\halfn,n;\halfn+1;-\frac{1}{\gamma_{i,j}}\right) -
                    \\
                    & \phantom{= \cdot\,\Biggl(}
                            \gamma_{i,j}^n\,\hgeo\left(\halfn,n;\halfn+1;-\gamma_{i,j} \right)
                        \Biggr) + 1
        \text{,}
    \end{split}
\end{equation}
where $\hgeo$ is the regularised Gauss hypergeometric function and $\gamma_{i,j}
= T_j / T_i$ is the ratio between the temperatures of two chains.  Since
$\E{A_{i,j}}$ depends on $T_i$ and $T_j$ only through the ratio $\gamma_{i,j}$,
uniform acceptance ratios between all adjacent pairs of chains can be achieved
with a geometric spacing of temperatures -- where $\gamma_{i,i+1}$ is constant
-- for a unimodal Gaussian likelihood.

\begin{figure}
    \begin{center}
        \includegraphics[width=\maxwidth{\columnwidth}]{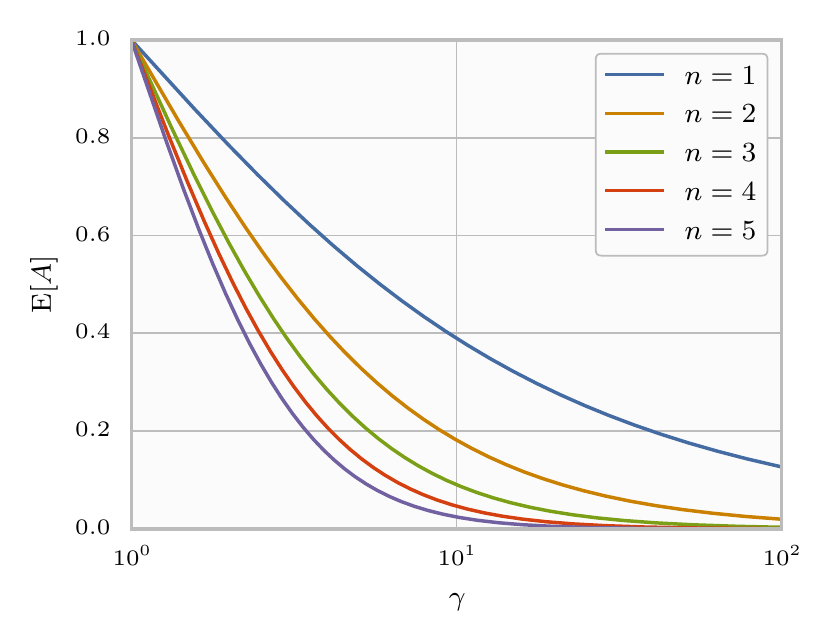}
        \caption[Acceptance ratio: ideal Gaussian]
        {\label{fig:ideal-gaussian-acceptance}
            The time-averaged acceptance ratio, $\E{A}$, between two chains
            of a PT\ac{MCMC} sampler on a unimodal, $n$-dimensional Gaussian
            likelihood distribution.  The chains have temperatures $T$ and
            $\gamma T$.
            The lines are ordered vertically to match the legend.
        }
    \end{center}
\end{figure}

The log spacing required for a particular acceptance ratio also depends on the
dimension of the parameter space, with more parameters requiring a closer
spacing of temperatures, illustrated by \autoref{fig:ideal-gaussian-acceptance}.
This can be understood by looking at the expectation and variance of $\logL$ at
a particular temperature (see \autoref{fig:ideal-gaussian-logl}),
\begin{equation}\label{eq:logl-exp-var}
    \E[T]{\logL} = -\frac{nT}{2}
    \quad\text{ and }\quad
    \Var[T]{\logL} = \frac{nT^2}{2}
    \text{.}
\end{equation}
Note that the specific heat from \eqref{eq:specific-heat} is a constant $n/2$,
as expected.

Since the acceptance ratio $A_{i,j}$ depends on $\logL_i - \logL_j$, the more
separate the distributions of $\logL_i$ and $\logL_j$ at their respective
temperatures, $T_i$ and $T_j$, the lower the acceptance ratio between such
chains will be.  For two chains at temperatures $T$ and $\gamma T$, the
separation of the means of $\ptilde_T$ and $\ptilde_{\gamma T}$, in units of the
standard deviation at $T$, will be
\begin{equation}\label{eq:logl-separation}
    \frac{\E[T]{\logL} - \E[\gamma T]{\logL}}
         {\sqrt{\Var[T]{\logL}}}
     = (\gamma-1)\sqrt{\halfn}
     \text{.}
\end{equation}

It follows that -- for constant $\gamma$ -- as the dimension $n$ increases, so
the acceptance ratio between chains at temperatures $T$ and $\gamma T$ falls.
For a higher dimensional target distribution, therefore, a closer spacing of
temperatures is required for a given acceptance ratio.

For more general distributions, by considering the overlap of $\ptilde_T(\logL)$
at different temperatures, \citet{Falcioni1999} argue that the number of
temperatures $N$ required to efficiently sample the posterior distribution
should scale with $\Delta\logL / \sqrt{n}$, where $\Delta\logL$ is the range of
$\E[T]{\logL}$ between $T = 1$ and $T = \Tprior$.  That is:
\begin{equation}\label{eq:chain-count-scaling}
    N \propto \frac{\E[1]{\logL} - \E[\Tprior]{\logL}}{\sqrt{n}}
    \text{.}
\end{equation}

Since the log likelihood range $\Delta\logL$ itself depends on the dimension of
the system $n$, it is difficult to apply this relation in practice.  However,
for the ideal Gaussian, we can see from \eqref{eq:logl-exp-var} that
$\Delta\logL$ scales with $n$, and so $N$ scales with $\sqrt{n}$, as we might
expect.

\subsection{The \acl*{KL} divergence}
\label{sec:kl}

Another measure of the optimal spacing of temperatures is the \acf{KL}
divergence between adjacent chains.  The \ac{KL} divergence from a hot
distribution $\pi_{T_j}$ to a cold distribution $\pi_{T_i}$,
\newcommand{\KL}{\ensuremath{D_\mathrm{KL}}}
\begin{equation}\label{eq:kl}
    \KL(\pi_{T_i}\|\pi_{T_j}) = \int \pi_{T_i}(\vectheta) \,
        \log\frac{\pi_{T_i}(\vectheta)}{\pi_{T_j}(\vectheta)} \,
        \dif\vectheta
    \text{,}
\end{equation}
quantifies the information gained about the posterior with each step down the
temperature ladder, from the prior $p = \pi_{T=\infty}$ to the posterior $\pi =
\pi_{T=1}$.  It is reasonable to expect that for an optimally-spaced ladder --
that is, one with a minimal \ac{ACT} on the cold chain for a given number of
chains -- the information gain should be uniform for every step down the ladder.

For the example of the ideal Gaussian of \autoref{sec:pt-ideal-gaussian}, the
\ac{KL} divergence is, straightforwardly,
\begin{equation}\label{eq:kl-gaussian}
    \KL(\pi_{T_i}\|\pi_{T_j}) = \halfn\left(\frac{1}{\gamma_{i,j}} + \log\gamma_{i,j} - 1\right)
    \text{.}
\end{equation}
Like the swap acceptance ratio, therefore, uniform \ac{KL} divergence over the
entire ladder is also achieved by a geometric spacing of temperatures for the
ideal Gaussian.

Unfortunately, unlike the acceptance ratio, the \ac{KL} divergence is difficult
to compute numerically while sampling, owing to the unknown -- and
temperature-dependent -- evidence (normalisation) values on $\pi_{T_i}$ and
$\pi_{T_j}$.

We henceforth assume that spacing temperatures for uniform acceptance ratios is
a reasonable approximation of a ladder that is optimal in the \ac{ACT} of the
cold chain.  We make this assumption on faith and, while we briefly examine its
validity in \autoref{sec:tests-gauss} and \autoref{sec:discussion-caveats}, it
invites a more careful study.

\section{Adaptive temperature ladders}
\label{sec:adaptive}

From the arguments in \autoref{sec:pt} and the references therein, we shall
assume that uniformity of acceptance ratios provides a good approximation to the
optimal temperature ladder for parallel tempering problems.  In this section, we
describe an algorithm for dynamically adapting chain temperatures to achieve
uniform acceptance ratios for inter-chain swaps.

From \eqref{eq:swap}, as $1/T_j - 1/T_i \to 0$, $A_{i,j} \to 1$, so in order to
increase the expected acceptance ratio between chains, it suffices to move them
closer together in temperature space; conversely, to reduce $\E{A_{i,j}}$, we
can push the chains apart.  We will henceforth adopt the notation that $A_i
\equiv A_{i,i-1}$ and that $T_i < T_{i+1}$, with $T_1 = 1$ being the untempered
or cold chain (which samples from the target distribution, $\pi$).  Here,
$A_i(t)$ are the instantaneous acceptance ratios between chains, but we shall
shortly describe the discrete case where empirical measurements of $A_i$ are
collected with each iteration of the sampler.

\subsection{Dynamics}
\label{sec:adaptive-dynamics}

Our goal is to dynamically adjust the temperatures of the chains to achieve
uniform acceptance ratios as we sample the target distribution.  We define our
temperature dynamics in terms of the log of the temperature difference between
chains,
\begin{equation}\label{eq:s-def}
    S_i \equiv \log(T_i - T_{i-1})
    \text{.}
\end{equation}
Under this scheme, finite changes to $S_i$ will always preserve the correct
ordering of temperatures ($T_1 < ... < T_N$).

To achieve the same $A_i$ for all chains, we can drive the gap $S_i$ according
to the acceptance ratios between chain $i$ and those immediately above and
below, to wit
\begin{equation}\label{eq:s-de}
    \od{S_i}{t} = \kappa(t) \left[A_i(t) - A_{i+1}(t)\right]
    \text{,}
\end{equation}
for $1 < i < N$, where $\kappa$ is a positive constant controlling the
time-scale of the evolution of $T_i$.  $\kappa$ can be interpreted as the
instantaneous exponential time-constant for temperature adjustments.  The two
extremal temperatures, $T_1$ and $T_N$, are fixed (see below).

Under this scheme, chain $i$ will attempt to increase the gap in temperature
space between itself and chain $(i+1)$ if swaps are accepted too often and close
it when they are accepted too seldom --- and similarly for chain $(i-1)$ ---
equilibrating at $A_i$ that are uniform over $i$.  Therefore, for an appropriate
choice of $\kappa$ -- discussed momentarily -- these rules drive the chains $i =
\{2,\ldots,N-1\}$ toward even acceptance spacing.

However, in order to efficiently sample a target distribution with strongly
separated modes (such that a traditional \ac{MCMC} sampler would be unable to
traverse the ``valleys'' between them), $T_{N}$ must be high enough that the
modes are flattened out and the chain can explore the entire parameter space
unhindered.  This amounts to the topmost chain sampling from the prior
distribution\footnote{
    For analytic priors, this special case, where the likelihood is ignored, can
    be treated separately by having the sampler draw independent samples
    directly from the prior.
}, which we achieve trivially by setting the inverse temperature of this chain
as $\beta_N = 0$.

This continuous system is discretised as
\begin{equation}\label{eq:s-de-discrete}
    S_i(t+1) - S_i(t) = \kappa(t)\left[A_i(t) - A_{i+1}(t)\right]
    \text{,}
\end{equation}
where $A_i(t)$ are the acceptance ratios accumulated by the sampler at the
current iteration.

The values of $A_i$ are measured empirically at each iteration as the fraction
of swap proposals between chains that were accepted.  For a traditional sampler
comprising one sample per chain, these will be either 0 or 1.  For ensemble
samplers, however, comprising $n_w$ distinct walkers per temperature, the
measurements of $A_i$ are less granular, such that $A_i \in
\{x\in[0,1]|n_wx\in\mathbb{Z}\}$.  In general, fewer walkers require a longer
averaging time-scale -- discussed below -- in order to smooth out this
granularity.

Importantly, the temperature adjustment scheme we have proposed -- and, more
generally, any adaptive sampling scheme -- in fact violates the condition for
detailed balance that ensures that an \ac{MCMC} sampler will converge to the
target distribution.  \citet{Roberts2007} investigate the conditions required of
such an adaptive sampler for it to be ergodic in the target distribution -- that
is, that it will converge on long time-scales.  They determine (from their
Theorem 1 and Corollary 4) that diminishing the amplitude of adaptations in the
transition kernel with each iteration is sufficient for the sampler to be
ergodic in the target distribution.  We therefore suppress temperature
adjustments to ensure that the sampler is Markovian on sufficiently long
time-scales\footnote{
    In principle, of course, we could stop temperature adjustments altogether
    once the temperatures have reached an equilibrium, discarding the previous
    samples as part of the burn-in.
}.

The rate of diminution of temperature adjustments is a trade-off between the
rate of convergence of the temperature ladder and that of the sampler itself
toward its stationary distribution.  We modulate the dynamics with hyperbolic
decay to suppress the dynamics on long time-scales,
\begin{equation}\label{eq:kappa}
    \kappa(t) = \frac{1}{\nu} \frac{t_0}{t+t_0}
    \text{,}
\end{equation}
where $t_0$ is the time at which the temperature adjustments have been reduced
to half their initial amplitude.  The initial amplitude of adjustments is in
turn set by $\nu$, the time-scale on which the temperatures evolve at early
time.

\subsection{Parameter choice}
\label{sec:adaptive-params}

In the scheme of \eqref{eq:s-de-discrete} and \eqref{eq:kappa}, there are two
parameters to choose: $t_0$ and $\nu$.  The dynamical time parameter $t$ in
\eqref{eq:s-de-discrete} is measured in units of intra-chain jumps of the
sampler, with temperature adjustments being made at every iteration.

The lag parameter $t_0$ sets the time-scale for the attenuation of temperature
adjustments.  This decay factor in $\kappa$ is included as a fail-safe mechanism
to ensure that, even for target distributions on which the temperature dynamics
fail to find an equilibrium set of temperatures, the ladder will always converge
over long time-scales.  This condition guarantees that the sampler correctly
explores the target distribution.

From \eqref{eq:kappa}, the time-scale of the dynamics at late time -- when $t
\gg t_0$ -- is $\nu t / t_0$.  To ensure that temperatures have time to find an
equilibrium over the course of a run, we therefore require that $t_0 \gg \nu$,
so that the dynamics will always be on a time-scale much shorter than the
current run time.  However, we should also ensure that, over the course of the
run, the dynamical time-scale is {\em longer} than the \ac{ACT} of the sampler,
so that the temperatures respond to the correct posterior distribution.  To this
end, we require that $\nu N_\tau \gg t_0$, where $N_\tau$ is the number of
independent samples gathered over the course of the run.  For example, if
$N_\tau = 100$, these two conditions are satisfied by $t_0 = 10\nu$, and for our
test cases, we have indeed found this choice to work well.

Meanwhile, the time-scale of the dynamics at early time -- when $t \ll t_0$ --
is $\nu$.  A good choice of $\nu$ should therefore ensure that the sampler is
not susceptible to large statistical errors on the measurements of the
acceptance ratios $A_i$.

In general, for $n_s$ swap proposals, the acceptance count $n_s A_i$ is a
random variable that follows a binomial distribution $B(n_s,\E{A_i})$, so that
$A_i$ has variance
\begin{equation*}
    \Var{A_i} = \frac{\E{A_i}(1 - \E{A_i})}{n_s}
    \text{.}
\end{equation*}
Since the dynamical equations \eqref{eq:s-de-discrete} are linear in $A_i$, they
will be driven by the means, $\E{A_i}$, on long time-scales, assuming that the
noise in the system from counting errors -- proportional to $1/\sqrt{n_s}$ --
does not cause short-term changes in $\E{A_i}$.

Given a sampler of $n_w$ walkers, $n_w$ swaps are proposed with each iteration,
so that $n_s = n_w\nu$.  To ensure stable dynamics at early time, we should
therefore choose $n_w\nu \gg 1$.

A good choice of $\nu$ depends on the response of $\E{A_i}$ to changes in the
relevant chains' temperatures, and therefore depends on the particular
likelihood function that is being sampled.  However, if $\E{A_i}$ will
eventually be of order, say, $0.25$, and we want the measurements of $A_i$ to be
between $0.2$ and $0.3$, then we should average $A_i$ over at least $100$ swap
proposals, giving $\nu \gtrsim 100/n_w$.

Combining these criteria on $\nu$ and $t_0$, we therefore suggest default
parameter values of $\nu = 10^2/n_w$ and $t_0 = 10^3/n_w$.

\section{Examples}
\label{sec:tests}

We have implemented the algorithm proposed above as a modification to the
ensemble sampler \emcee{} of \citet{emcee}.  Our implementation can be found at
\emceeref.

In this section we apply our implementation to specific examples in order to
understand how and when the traditional geometric spacing fails and how much the
uniform-$A$ strategy might help us.  We present the following test cases.
\begin{enumerate}
    \item In \autoref{sec:tests-gauss} we compare the uniform-$A$ strategy used
        by the temperature dynamics of \autoref{sec:adaptive} with the
        alternative strategy of uniform \ac{KL} divergence discussed in
        \autoref{sec:pt-ideal-gaussian} on the example of a unimodal truncated
        Gaussian likelihood.
    \item In \autoref{sec:tests-2banana} we test the dynamics on a more complex,
        bimodal distribution for various choices of the number of chains $N$.
        We compare the resulting \acp{ACT} of the sampler with those of another
        sampler using a geometric ladder whose maximum temperature is fixed such
        that $\Tmax \approx \Tprior$.
    \item In \autoref{sec:tests-eggbox} we test the algorithm against the more
        difficult egg-box distribution with 243 modes.  For comparison, we
        sample from the same distribution with a geometric ladder constructed to
        yield 25\% acceptance ratios when applied to the ideal Gaussian
        discussed in \autoref{sec:pt-ideal-gaussian}.
\end{enumerate}

For all of these tests, $\nu = 10^2$ and $t_0 = 10^3$ are used to control the
dynamics in \eqref{eq:kappa}, while the sampler uses $100$ walkers.  Note that
these choices, while different from the defaults proposed in
\autoref{sec:adaptive-params}, do satisfy the conditions described in that
section.

\subsection{Truncated Gaussian}
\label{sec:tests-gauss}

Our first test case is an $n$-dimensional, unimodal, unit Gaussian similar to
that of \autoref{sec:pt-ideal-gaussian} but with finite prior volume.  The
simplicity of this case admits some exact analysis before recourse to numerics,
which allows us to test the approximations made in
\autoref{sec:pt-ideal-gaussian}.

At low temperatures, where the prior boundaries do not truncate much of the
likelihood probability mass, the optimal temperature spacing should be similar
to that of the ideal Gaussian.  By imposing a step-like cut-off in the prior at
a radius of $R$, there will be some temperature at which this approximation will
fail and a geometric spacing becomes inappropriate.

For the likelihood we use the same distribution as in
\autoref{sec:pt-ideal-gaussian}, while for the prior we use a uniform
distribution over the closed $n$-ball of radius $R = 30$, centred on the origin.
The likelihood and prior are defined by
\begin{align}
    \label{eq:gauss-likelihood}
    L(\vectheta) &\propto \exp\left(-\half\,\vecthetasq\right)
    \text{,} \\
    \label{eq:gauss-prior}
    p(\vectheta) &\propto \begin{cases}
                            1 & \text{ if } \|\vectheta\| \le R \text{,} \\
                            0 & \text{ otherwise,}
                          \end{cases}
\end{align}
where $\|\cdot\|$ is the Euclidean norm on $\mathbb{R}^n$.  Subsequently, the
normalised posterior generated by \eqref{eq:gauss-likelihood} and
\eqref{eq:gauss-prior} at temperature $T$ is
\begin{equation}\label{eq:gauss-posterior}
    \pi_T(\vectheta) =
                        \begin{cases}
                            \frac
                            {\left(2\pi T\right)^{-\halfn} \Gamma\left(\halfn\right)}
                                {\tilde{\gamma} \left(\halfn, \frac{R^2}{2T}\right)}
                            \exp\left(-\frac{\vecthetasq}{2T}\right)
                                & \text{ if } \|\vectheta\| \le R \text{,} \\
                            0
                                & \text{ otherwise,}
                        \end{cases}
\end{equation}
where $\tilde{\gamma}(a,z)$ is the lower incomplete gamma function.

In the low-temperature limit, this distribution converges to the ideal Gaussian
distribution.  We should therefore expect the \ac{KL} divergence for a step down
the temperature ladder to asymptote to \eqref{eq:kl-gaussian} as $T \to 0$,
where the effects of the prior boundary are negligible\footnote{
    While we do not consider $T < 1$ in our simulations, the case of $T \to 0$
    can equivalently be thought of as $R \to \infty$, since the width of the
    Gaussian scales with $\sqrt{T}$.
}.  Indeed, the \ac{KL} divergence of \eqref{eq:gauss-posterior} from $T_2$ to
$T_1$ is available analytically as
\begin{equation}\label{eq:kl-gauss}
    \begin{split}
        \KL = & -\frac{(T_2 - T_1) \, \tilde{\gamma}\left(1 + \halfn, \frac{R^2}{2 T_2}\right)}
                    {T_2 \, \tilde{\gamma}\left(\halfn, \frac{R^2}{2 T_1}\right)}
              \\
              & +
              \halfn\log\left( \frac{T_2}{T_1} \right)
              +
              \log\left(
                  \frac{\tilde{\gamma}\left( \halfn, \frac{R^2}{2 T_2} \right)}
                       {\tilde{\gamma}\left( \halfn, \frac{R^2}{2 T_1} \right)}
              \right)
        \text{.}
    \end{split}
\end{equation}
%
%\ilya{[Didn't check last two equations -- hopefully, someone will]}
%
If we set $T_2 = \gamma T_1$ (with $\gamma T_1 \ll 1$), then
$\tilde{\gamma}(a,z) \to \Gamma(a)$ as $T_1 \to 0$, and the expression reduces
to \eqref{eq:kl-gaussian}, as expected.

\begin{figure}
    \begin{center}
        \includegraphics[width=\maxwidth{\columnwidth}]{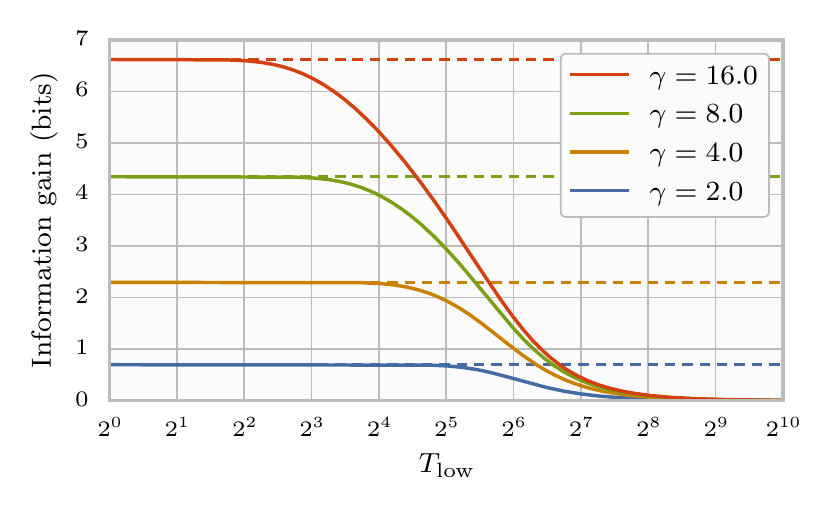}
        \caption[KL divergence: truncated Gaussian]
        {\label{fig:kl-gauss}
            The \ac{KL} divergence, or information gain, from a hot chain at
            temperature $\gamma \, T_\mathrm{low}$ to a colder chain at
            temperature $T_\mathrm{low}$, both sampling from
            \eqref{eq:gauss-posterior} at $n = 5$ (solid lines).  As
            $T_\mathrm{low} \to 0$, the information gain tends to that of the
            ideal Gaussian of \autoref{sec:pt-ideal-gaussian} (dashed lines).
            The lines are ordered vertically to match the legend.
        }
    \end{center}
\end{figure}

\autoref{fig:kl-gauss} illustrates this convergence for $n = 5$.  The point on
this plot at which the solid line diverges from the dashed line, for each
$\gamma$, predicts the temperature beyond which a geometric spacing of
temperatures is no longer optimal (for optimality as defined by uniform \ac{KL}
divergence between chains).  This is caused by truncation of the tempered
likelihood by the prior boundaries.

Of course, since the \ac{KL} divergence cannot easily be assessed empirically by
an \ac{MCMC} sampler, and we must instead resort to using acceptance ratios, we
would like to know how consistent these two schemes are outside the assumptions
of \autoref{sec:pt-ideal-gaussian}.

\begin{figure*}
    \begin{center}
        \includegraphics[width=\maxwidth{\textwidth}]{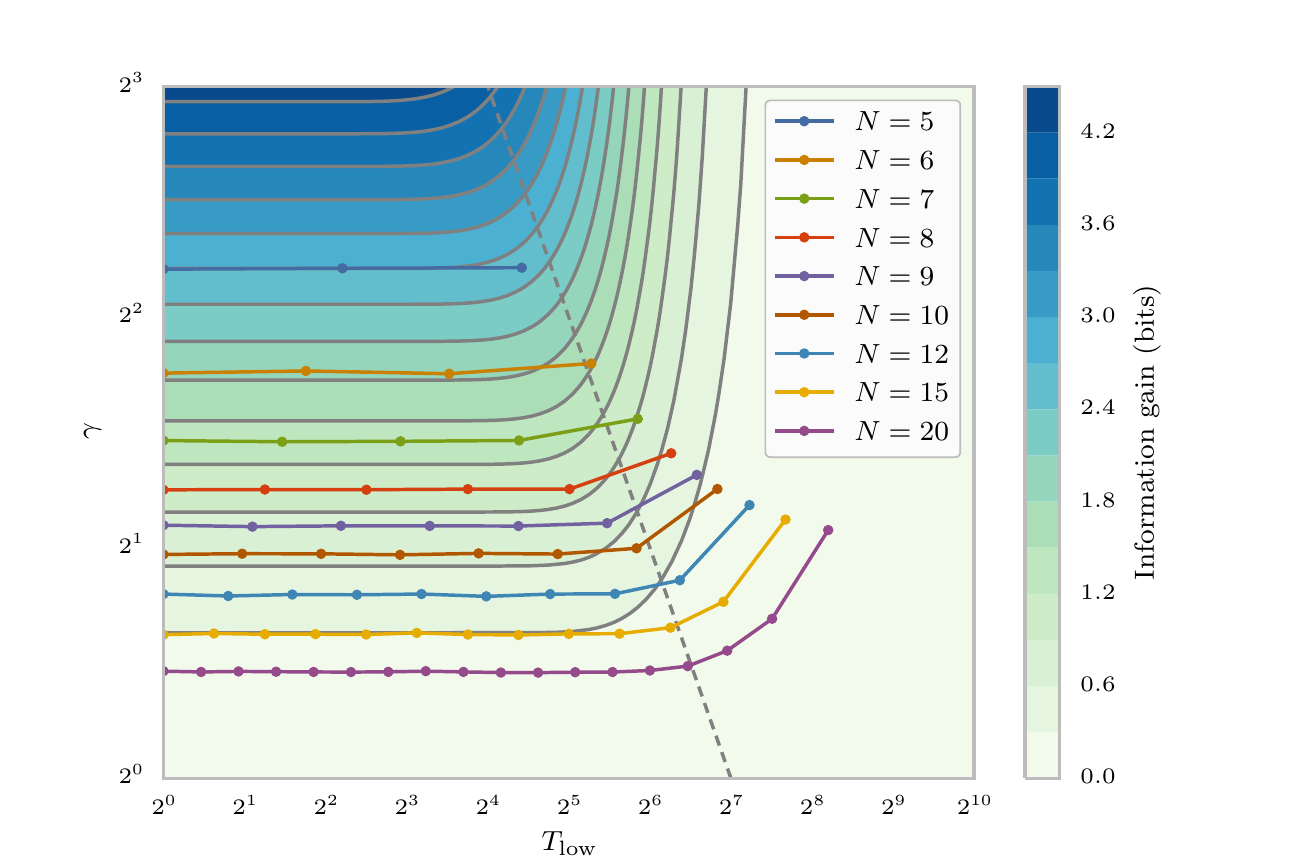}
        \caption[KL divergence vs. acceptance ratio]
        {\label{fig:kl-gauss-contour}
            A contour plot of the \ac{KL} divergence, or information gain, from
            a hot chain at temperature $T_\mathrm{high} =
            \gamma\,T_\mathrm{low}$ to a colder chain at temperature
            $T_\mathrm{low}$, both sampling from the Gaussian likelihood
            \eqref{eq:gauss-posterior}.  The coloured lines show the equilibrium
            $N$-chain temperature ladders reached by the temperature dynamics
            algorithm of \autoref{sec:adaptive}, where the acceptance ratio is
            the same between any pair of adjacent chains.  The points on these
            lines represent pairs of adjacent temperatures $(T,\gamma T)$
            (excluding the top-most, where $\gamma = \infty$).
        }
    \end{center}
\end{figure*}

\autoref{fig:kl-gauss-contour} shows contours of constant $\KL$, calculated from
\eqref{eq:kl-gauss}, and contours of constant $A_i$, illustrated by points
representing temperature pairs (from ladders selected by the algorithm developed
in \autoref{sec:adaptive}).  In the low temperature limit, as expected, both
schemes select a geometric spacing of temperatures consistent with the ideal
Gaussian of \autoref{sec:pt-ideal-gaussian} (i.e., the contours remain constant
in $\gamma$).  At higher temperatures, both schemes depart from the geometric
spacing, but they do so differently.  The uniform acceptance scheme displays a
more gradual departure from a geometric spacing than the contours of constant
$\KL$.  The smaller $\gamma$ selected by the uniform-$A$ scheme outside the
geometric regime, however, suggest that closer spacing is required in difficult
temperature ranges (e.g., across a phase transition) in order to achieve uniform
$A$ than would be required for uniform $\KL$.  There is therefore less risk of a
large gap in temperature across such a temperature range, at the cost of
(potentially) slightly less efficient communication across the rest of the
ladder.  The uniform-$A$ criterion for optimality is therefore conservative with
respect to a uniform-$\KL$ criterion.

We can also visualise the ladder specification in terms of the density of chains
over temperature.  We define this density, in $\log T$, as
\begin{equation}\label{eq:eta}
    \eta(\log T) = \od{N}{\,\log T} = \frac{1}{\Delta\log T} = \frac{1}{\log\gamma}
    \text{,}
\end{equation}
with $\gamma = T_{i+1}/T_i$, where $T_{i+1}$ and $T_i$ are the chain
temperatures to either side of $T$.

\begin{figure}
    \begin{center}
        \includegraphics[width=\maxwidth{\columnwidth}]{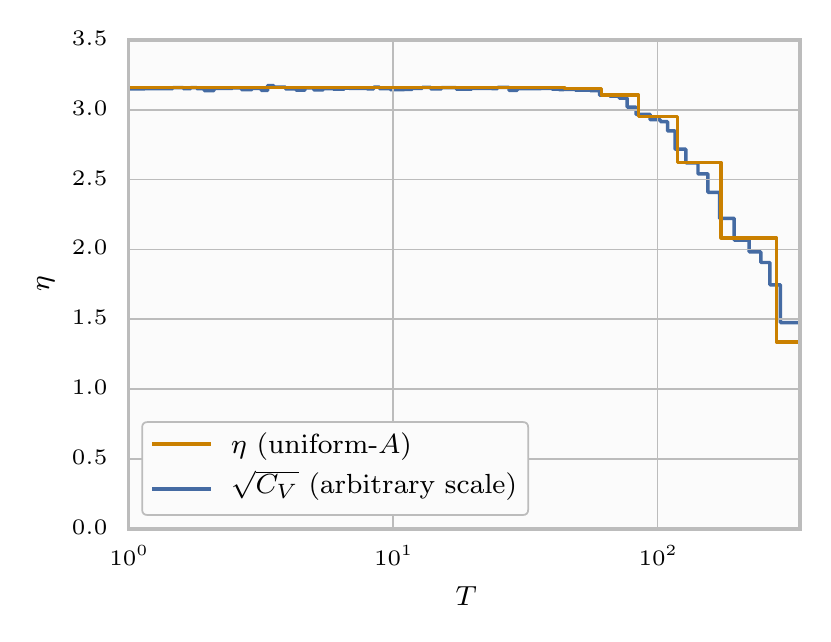}
        \caption[Chain density: truncated Gaussian]
        {\label{fig:gauss-25-eta-heat}
            {\bf Orange:} The density of chains per $\log T$ under the truncated
            Gaussian distribution \eqref{eq:gauss-posterior}, where $N = 20$, $n
            = 25$, and temperatures are chosen for uniform acceptance ratios
            between chains.  The chains have equilibrated to 77\% acceptance.
            {\bf Blue:} The square root of the specific heat of the truncated
            Gaussian distribution, normalised to match the chain density $\eta$
            of the uniform-$A$ ladder, between $T_1$ and $T_{N-1}$.  The
            specific heat $C_V$, from \eqref{eq:specific-heat}, is estimated
            from the sample means of $\logL$ over many runs with different
            temperature ladders.
        }
    \end{center}
\end{figure}

\autoref{fig:gauss-25-eta-heat} shows this density for a temperature ladder of
20 chains that is in equilibrium under the temperature dynamics of
\autoref{sec:adaptive} (the $N = 20$ contour of \autoref{fig:kl-gauss-contour}).
The density exhibits the expected uniformity of $\gamma$ for low temperatures
but falls for $T \gtrsim 80$.  The width $\sigma$ of the unit Gaussian at
temperature $T$ is $\sqrt{T}$, so at this temperature the prior boundary is at
$\sim 3\sigma$.  At $T = 80$, $\sim 5\%$ of the likelihood mass is truncated --
compared to $< 0.1\%$ for $T = 40$ and $\sim 35\%$ for $T = 160$ -- indicating
that the prior boundary becomes significant in this temperature regime.

This drop in density reflects the convergence of the tempered posterior
distribution, $\pi_T$, toward the prior as $T \to \infty$.  As $\pi_T$ becomes
flatter, fewer chains are needed per $\log T$ to maintain good communication.

Also shown on figure \autoref{fig:gauss-25-eta-heat} is the square root of the
estimated specific heat $C_V$ of the system as discussed in
\autoref{sec:pt-ladder-selection}, which can be seen to track closely the
logarithmic chain density $\eta$ when appropriately normalised.  While the
provenance of this relationship is unclear, it demonstrates the relevance of the
specific heat in determining an effective temperature ladder.

\subsection{Double Rosenbrock function}
\label{sec:tests-2banana}

The previous test demonstrated how a geometric ladder spaces temperatures too
closely at higher temperatures, as the prior boundary becomes significant.
While this may be an inefficient use of resources, it at least doesn't
drastically inhibit communication between high temperatures and low
temperatures.  Instead, we now turn to a more complex, bimodal likelihood
distribution for which a geometric spacing might cause bottlenecks in the
communication between high and low temperatures.

We use a likelihood derived from the two-dimensional Rosenbrock function $f$:
\begin{equation}\label{eq:2banana-likelihood}
    L(x,y) \propto \left( \frac{1}{c + f(x,y)} + \frac{1}{c + f(-x,y)} \right)^{1/T_p}
    \text{,}
\end{equation}
where
\begin{equation}\label{eq:rosenbrock}
    f(x,y) = (a - x)^2 + b(y - x^2)^2
    \text{.}
\end{equation}

$T_p$ is a pre-tempering factor chosen to increase the contrast of the
distribution, making it harder to sample.  When $T_p \ll 1$, each mode is
locally Gaussian, making the results comparable to the Gaussian example
considered in \autoref{sec:pt-ideal-gaussian}.

\begin{figure}
    \begin{center}
        \includegraphics[width=\maxwidth{\columnwidth}]{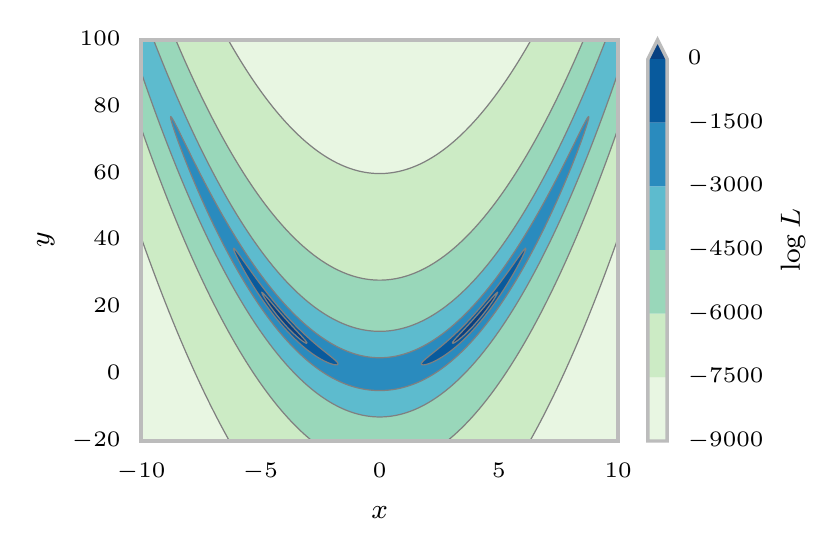}
        \caption[Double Rosenbrock log likelihood]
        {\label{fig:2banana-likelihood}
            The Rosenbrock log likelihood, from \eqref{eq:2banana-likelihood}.
        }
    \end{center}
\end{figure}

For the following tests, we use $a = 4$, $b = 1$, $c = 0.1$, and $T_p =
10^{-3}$.  We use a flat prior on $[-10,10]\times[-20,100]$.
\autoref{fig:2banana-likelihood} illustrates this likelihood over the prior
volume.

\begin{figure*}
    \begin{center}
        \includegraphics[width=\maxwidth{\textwidth}]{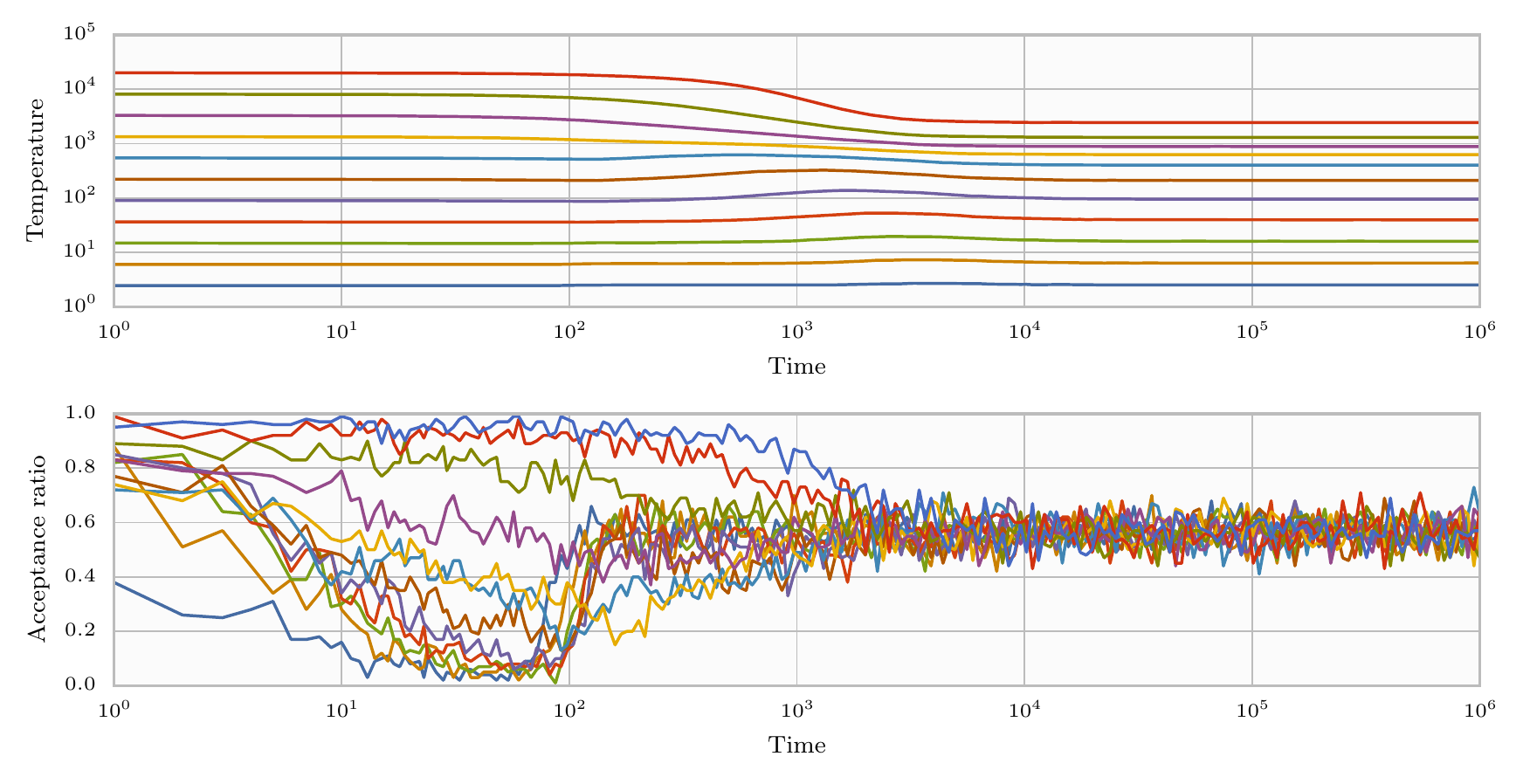}
        \caption[Temperature evolution: double Rosenbrock]
        {\label{fig:2banana-evolution}
            The evolution of ladder of 13 temperatures $T_i$ and acceptance
            ratios $A_i$ over an \emcee{} run of $10^6$ iterations under the
            Rosenbrock likelihood \eqref{eq:2banana-likelihood}.  Chains 1 and
            13 are not shown, having fixed temperatures $T_1 = 1$ and $T_{13} =
            \infty$.
        }
    \end{center}
\end{figure*}

\begin{figure}
    \begin{center}
        \includegraphics[width=\maxwidth{\columnwidth}]{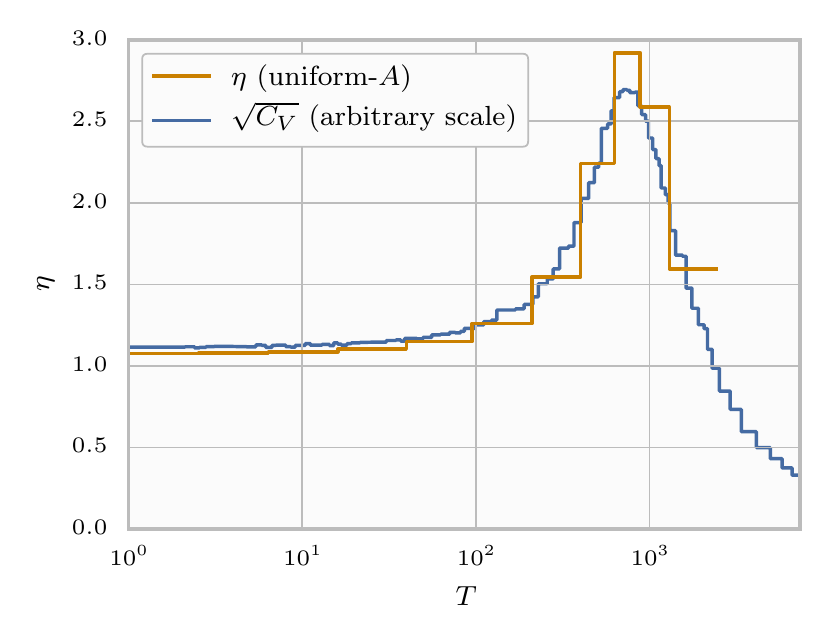}
        \caption[Chain density: double Rosenbrock]
        {\label{fig:2banana-eta-heat}
            {\bf Orange:} The equilibrium density of chains per $\log T$ for the
            double Rosenbrock run illustrated in \autoref{fig:2banana-evolution},
            where the acceptance ratios have settled to $A_i \approx 0.57$.
            {\bf Blue:} The square root of the specific heat for the double
            Rosenbrock distribution, as described in
            \autoref{fig:gauss-25-eta-heat}.
        }
    \end{center}
\end{figure}

\subsubsection{Test: temperature evolution}
\label{sec:tests-2banana-evolution}

As an illustrative example, we first tested the temperature dynamics of
\autoref{sec:adaptive} with the double Rosenbrock posterior distribution
\eqref{eq:2banana-likelihood} using 13 chains.  \autoref{fig:2banana-evolution}
shows the evolution of the temperature ladder according to these dynamics, while
\autoref{fig:2banana-eta-heat} shows the chain density $\eta(\log T)$ for the
equilibrated temperature ladder.

While the equilibrated chains are distributed uniformly in $\log T$ for $T
\lesssim 50$, there is a distinct peak in $\eta$ at $T \approx 800$, where a
simple geometric spacing of temperatures hinders communication between chains.
This peak occurs at a phase transition where the two modes of the likelihood
distribution begin to mix and $\E{\logL}$ changes rapidly with $T$, indicated by
the sharp change in specific heat in the bottom panel of
\autoref{fig:2banana-eta-heat}.  Since the shape of the likelihood distribution
in this regime becomes very sensitive to $T$, a higher density of chains is
needed to maintain a given acceptance ratio.  We also note that in the geometric
regime (i.e., for low $T$) the specific heat is approximately $n/2 = 1$, with
$\E{A} \approx 57\%$, consistent with the values derived for the ideal Gaussian
from \eqref{eq:logl-exp-var} and \eqref{eq:acceptance-solution} respectively.

Ultimately, however, the figure of merit for a temperature specification in a
PTMCMC simulation is the resulting \ac{ACT} for the target temperature ($T = 1$)
of the sampler.  We must therefore test the performance of the sampler
empirically.

We use the term \ac{ACT} to refer to the {\em integrated autocorrelation time}
discussed by \citet{Sokal1997}, which we estimate according to the algorithm
used in the {\em acor} package (see \appref{sec:act-estimation} and
\url{http://www.math.nyu.edu/faculty/goodman/software/acor/} for details).  For
the following tests, we use the \ac{ACT} of the first parameter, $x$, as a
measure of the efficiency of the sampler (since \eqref{eq:2banana-likelihood} is
bimodal in $x$ but unimodal in $y$).

\subsubsection{Test: improvement over a geometric ladder}
\label{sec:tests-2banana-improvement}

In \autoref{sec:pt} we claimed that aiming for uniform acceptance ratios between
chains yields a good temperature ladder.  Specifically, we expect that a ladder
selected for uniform acceptance ratios should lead to a lower \ac{ACT} for the
$T = 1$ chain than that resulting from a plain geometric ladder.

\begin{figure}
    \begin{center}
        \includegraphics[width=\maxwidth{\columnwidth}]{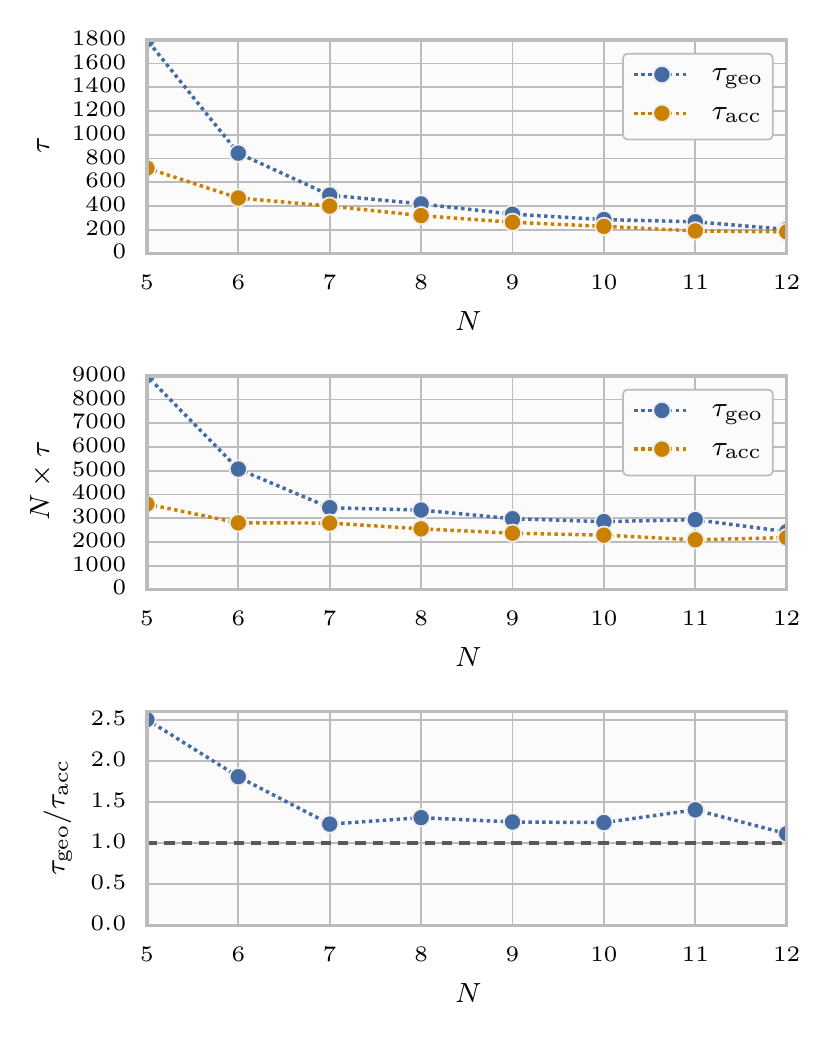}
        \caption[Autocorrelation times: double Rosenbrock]
        {\label{fig:2banana-act}
            {\bf Top:} the \acp{ACT} of $x$ for the cold ($T = 1$) chain of a
            sampler exploring the double Rosenbrock distribution
            \eqref{eq:2banana-likelihood}, using both uniform-$A$ and
            geometrically spaced temperature ladders as a function of the number
            of chains $N$.
            {\bf Middle:} the total CPU time, $N \times \tau$, for the runs.
            {\bf Bottom:} the relative improvement in the \ac{ACT} for the
            uniform-$A$ ladder over the geometric ladder.  The joining lines are
            provided to guide the eye.
        }
    \end{center}
\end{figure}

The geometric ansatz that we use has a fixed maximum temperature such that $T_N
= 2\times10^4$.  As $N$ increases, more chains are added between $T_1$ and
$T_N$, maintaining the geometric spacing.  Under this arrangement, the addition
of new temperatures is not redundant even when $T_N$ is already high enough to
sample from the prior; the additional chains instead aid inter-chain
communication at lower temperatures.  Since $T_N$ is close to the temperature at
which the posterior becomes the prior, there is little CPU time wasted in
sampling redundantly from the prior with several chains, while lower-temperature
chains can still communicate with a chain sampling from the prior.  Under this
set-up, therefore, the \ac{ACT} always decreases as $N$ increases, per
\autoref{fig:2banana-act}.

To test the improvement in \ac{ACT}, $\tau$, conferred by our temperature
dynamics, we allowed \emcee{} to explore the target distribution
\eqref{eq:2banana-likelihood} with different numbers of chains, $N$, using both
the uniform-$A$ ladders and geometrically spaced ladders.  The resulting
\acp{ACT}, $\actgeo$ and $\actacc$, are plotted against $N$ in
\autoref{fig:2banana-act}.

In this example, an $N$-chain ladder dynamically adapted for uniform acceptance
ratios clearly outperforms a geometrically spaced ladder of the same size for
all $N$.

The benefit of a uniform-$A$ ladder is most pronounced at low $N$ -- i.e., where
there are few chains available.  In this regime, the sampler will be more
sensitive to phase transitions, since the bigger gaps in temperature could cause
severe bottlenecks in communication across the temperature ladder.

When $N$ is large, the differences in acceptance ratios between a geometric
ladder and one chosen for uniform $A$ becomes less significant.  In this case,
the difference between the limiting (minimum) acceptance ratio for a ladder and
the ladder's average acceptance ratio is proportionally smaller.

In the case of the double Rosenbrock distribution \eqref{eq:2banana-likelihood},
we have found that, once the minimum acceptance ratio for a geometric ladder
(terminating at $\Tmax = 2\times10^4$) exceeds $\sim 10\%$, reallocating
temperatures for uniform acceptance ratios does not reduce the measured \ac{ACT}
by more than $25\%$.  This occurs when $N \approx 7$ in the current example.
Nonetheless, there remains an overall improvement in \ac{ACT} regardless of $N$.

\autoref{fig:2banana-act} also shows, in the middle pane, the total number of
iterations per independent sample across all chains.  This quantity, given by $N
\times \tau$, is proportional to the total CPU time of the simulation, while
$\tau$ itself is proportional to the CPU time per chain, or {\em wall time}, of
the simulation.  In this instance, the CPU time of a run diminishes with $N$ in
much the same fashion as the wall time does.  The fractional improvement in CPU
time is of course the same as for wall time -- $\actgeo/\actacc$.

\subsubsection{Test: chain removal}
\label{sec:tests-2banana-removal}

To determine whether a uniform-$A$ temperature placement strategy is in fact
close to optimal, we assess the contribution of each chain from such a
temperature ladder to the efficiency of the sampler, as measured by its
\ac{ACT}.  If this contribution is equal for all chains, then we can conclude
that it is indeed optimal to have them all exchanging equally -- that is, with
uniform acceptance ratios.

To this end, we conducted the following test:
\begin{enumerate}
    \item Sample from \eqref{eq:2banana-likelihood} with $N = 7$ chains under
        the temperature dynamics of \autoref{sec:adaptive} until the temperatures
        have equilibrated to $(T_1,\ldots,T_7)$ to give uniform acceptance
        ratios.
    \item Generate 5 new test ladders, each of 6 chains, formed by removing the
        $i^\mathrm{th}$ chain from that determined above -- i.e.,
        $(T_1,\ldots,T_{i-1},T_{i+1},\ldots,T_7)$ -- for $i = 2,\ldots,6$.
    \item Sample from \eqref{eq:2banana-likelihood} with each of these 5 test
        ladders and calculate the \acp{ACT} on the cold chain, $\acttest$.
\end{enumerate}

\begin{figure}
    \begin{center}
        \includegraphics[width=\maxwidth{\columnwidth}]{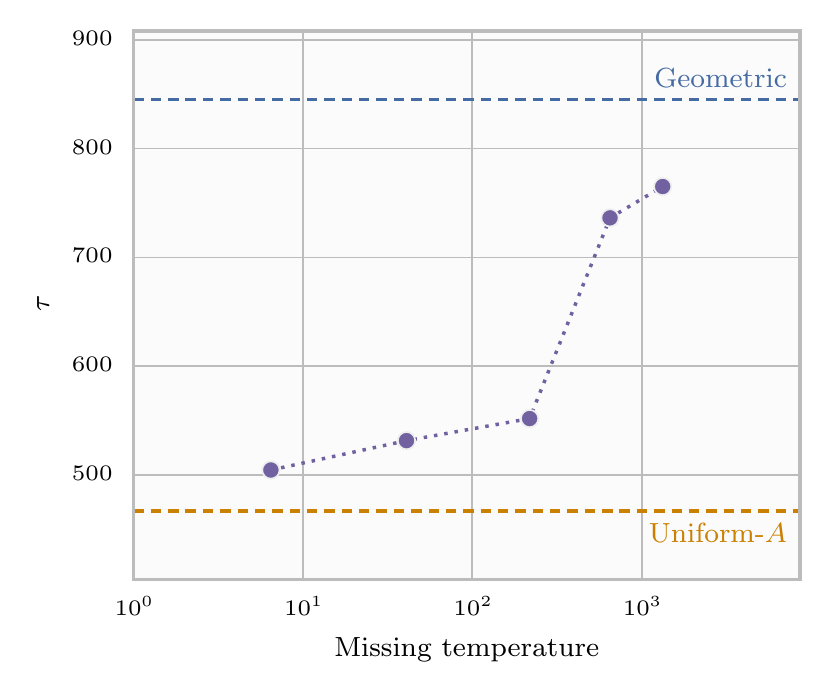}
        \caption[Autocorrelation times: chain removal test]
        {\label{fig:2banana-removal-act}
            The cold-chain \acp{ACT} for samplers exploring the double
            Rosenbrock distribution \eqref{eq:2banana-likelihood} per the test
            described in \autoref{sec:tests-2banana-removal}.  The points denote
            the \acp{ACT} from ladders generated according to the scheme in
            \autoref{sec:tests-2banana-removal}.  The dashed lines above and
            below identify the \acp{ACT} from geometric and uniform-$A$ ladders,
            respectively, of $N = 6$.
        }
    \end{center}
\end{figure}

\begin{figure}
    \begin{center}
        \includegraphics[width=\maxwidth{\columnwidth}]{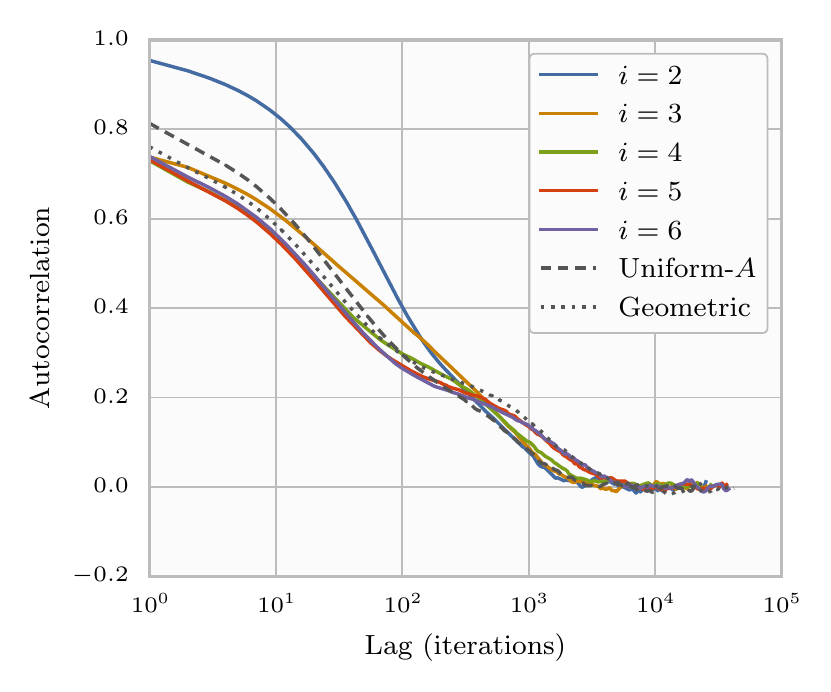}
        \caption[Autocorrelation function: double Rosenbrock]
        {\label{fig:2banana-removal-acf}
            The cold-chain autocorrelation function for a sampler exploring the
            double Rosenbrock distribution \eqref{eq:2banana-likelihood}.  The
            solid lines correspond to the ladders generated by the scheme
            outlined in \autoref{sec:tests-2banana-removal}, where $i$ is the
            index of the removed chain.  For comparison, the dashed and dotted
            lines represent respectively uniform-$A$ and geometric ladders of
            the same size.  The approximate \acp{ACT} are 504, 531, 551, 736,
            and 765, for $i = 2,\ldots,6$; 467 for a uniform-$A$ ladder; and 844
            for a geometric ladder (see \autoref{fig:2banana-removal-act}).
        }
    \end{center}
\end{figure}

\autoref{fig:2banana-removal-act} shows the \acp{ACT} for the cold chain
resulting from the test outlined above.  While $\actacc < \acttest < \actgeo$
for ladders of the same $N$, $\acttest$ increases with the temperature of the
chain that is removed, suggesting that additional chains are more useful at
higher temperatures.  The sharp jump in $\acttest$ when a chain above $T \approx
200$ is removed arises from the phase transition that occurs as $T$ approaches
$\Tprior$, indicated by a peak in $C_V$ (visible in
\autoref{fig:2banana-eta-heat}).

We can understand this behaviour by examining the complete autocorrelation
functions from which these \acp{ACT} are estimated.  Illustrated in
\autoref{fig:2banana-removal-acf}, these autocorrelation functions exhibit two
distinct time-scales.  Firstly, there is a large autocorrelation for lags
$\lesssim 100$ for all $i$ -- particularly $i = 2$ -- corresponding to the
\ac{ACT} of the sampler within one of the two modes: that is, the time taken for
the sampler to generate an independent sample without changing mode.  Secondly,
there is a visible hump in the autocorrelation function for $100 \lesssim$ lag
$\lesssim 2000$, corresponding to the time taken for the sampler to migrate
between modes.  Removing the second chain from initial geometric ladder of 7
chains increases the intra-mode \ac{ACT} in particular, but does not affect the
inter-mode \ac{ACT}.  Meanwhile, while removing higher temperature chains pushes
the secondary hump outward to larger lags, increasing the inter-mode \ac{ACT}
instead.

The overall autocorrelation time in which we are interested, discussed by
\citet{Sokal1997} and in \appref{sec:act-estimation}, represents the time
between independent samples of the system.  It is therefore set by the
time-scale on which the sampler migrates to a new mode independently of the
current mode.  Removing a chain at higher temperatures increases the inter-modal
\ac{ACT}, and therefore damages the efficiency of the sampler.

Nonetheless, all of the tested temperature ladders yielded lower \ac{ACT} than
the default geometric ladder, despite the geometric ladder being chosen with
prior knowledge of $\Tprior$.

\subsection{Egg-box in five dimensions}
\label{sec:tests-eggbox}

To test the algorithm's performance on a yet more strongly multi-modal
distribution, we use an egg-box distribution defined by the likelihood
\begin{equation}\label{eq:eggbox-5-likelihood}
    L(\vectheta) \propto \left( \half\prod_{i=1}^n \cos\theta_i + \half \right)^{1/T_p}
    \text{.}
\end{equation}
For a small value of the pre-tempering factor $T_p$ the modes of this
distribution become locally Gaussian, and in the low-$T$ regime should therefore
generate results similar to those of the Gaussian distributions examined in
\autoref{sec:pt-ideal-gaussian} and \autoref{sec:tests-gauss}.  For the following
tests, we choose $T_p = 10^{-3}$.

We explore this likelihood distribution in 5 dimensions over a flat prior on
$[-L/2,L/2]^n$, where we choose $L = 3\pi$, giving $3^n = 243$ modes.

Rather than compare our uniform-$A$ temperature ladder against a geometric
ladder with a fixed maximum temperature, as in \autoref{sec:tests-2banana}, we
instead use a geometric ladder constructed to give a fixed acceptance ratio of
$\E{A} = 0.25$ when applied to the special case of an ideal Gaussian likelihood
(per \autoref{sec:pt-ideal-gaussian}).  Such a ladder will not, in general, give
uniform acceptance ratios when applied to an arbitrary posterior distribution,
but this choice reflects the more realistic scenario where we cannot guess at
$\Tprior$, and so we resort to assuming that the distribution indeed behaves
like an ideal Gaussian.

\begin{figure*}
    \begin{center}
        \includegraphics[width=\maxwidth{\textwidth}]{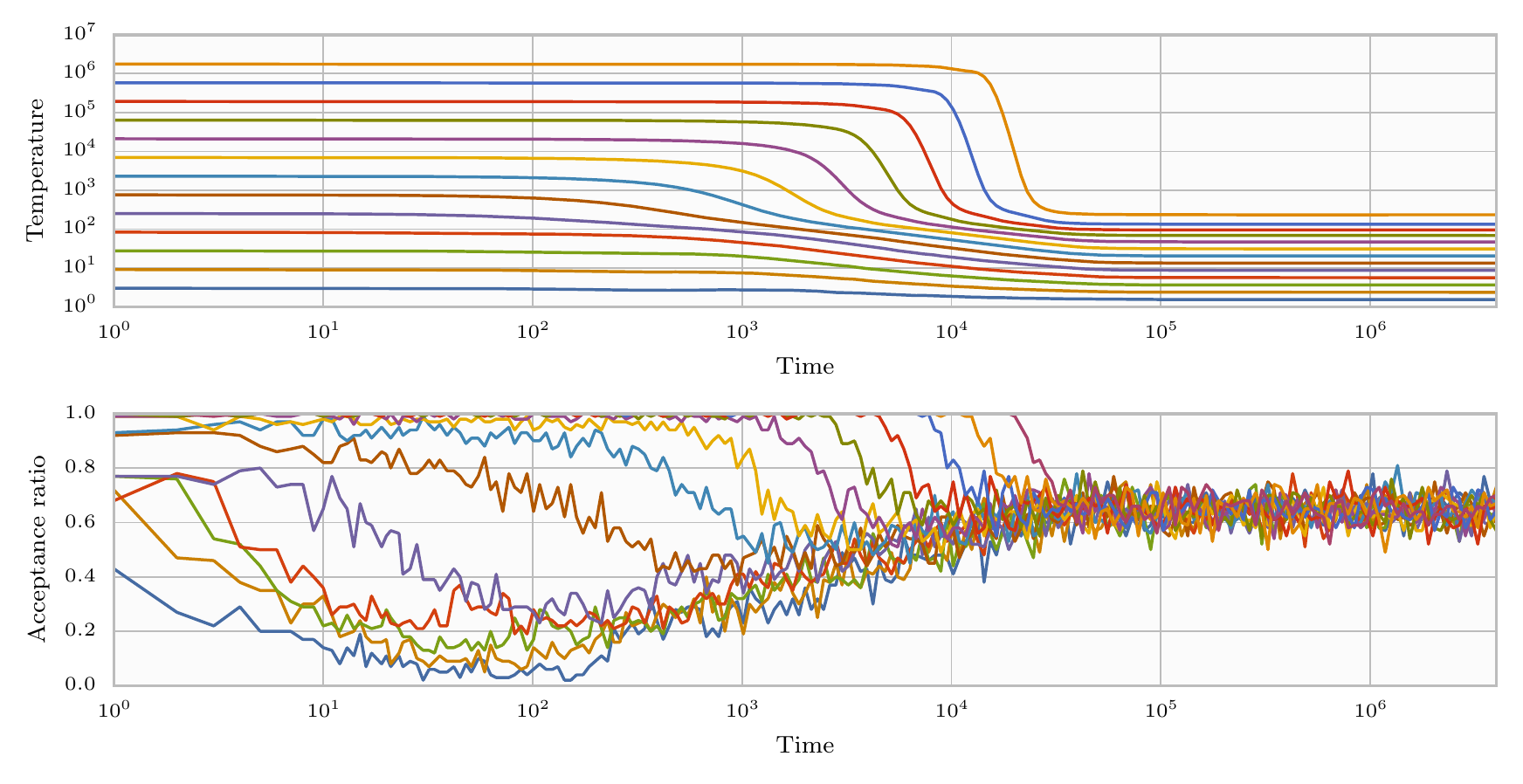}
        \caption[Temperature evolution: eggbox]
        {\label{fig:eggbox-5-evolution}
            The evolution of temperatures $T_i$ and acceptance ratios
            $A_i$ while sampling with \emcee{} from a 5-dimensional egg-box
            distribution, \eqref{eq:eggbox-5-likelihood}, with 15 chains.
            Chains 1 and 15 are not shown, having fixed temperatures $T_1 = 1$
            and $T_{15} = \infty$.
        }
    \end{center}
\end{figure*}

\begin{figure}
    \begin{center}
        \includegraphics[width=\maxwidth{\columnwidth}]{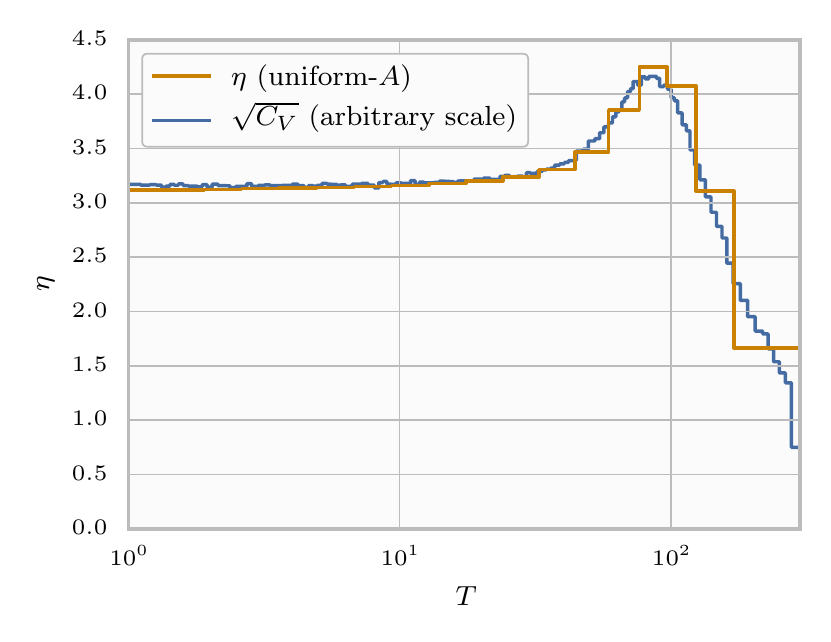}
        \caption[Chain density: eggbox]
        {\label{fig:eggbox-5-eta-heat}
            {\bf Orange:} The equilibrium density of chains per $\log T$ for the
            egg-box run illustrated in \autoref{fig:eggbox-5-evolution}, where
            the acceptance ratios have settled to $A_i \approx 0.65$.
            {\bf Blue:} The square root of the specific heat for the egg-box
            distribution, as described in \autoref{fig:gauss-25-eta-heat}.
        }
    \end{center}
\end{figure}

\autoref{fig:eggbox-5-evolution} shows the evolution of the temperatures and
acceptance ratios for an \emcee{} sampler of 15 chains under the temperature
dynamics of \autoref{sec:adaptive}.  \autoref{fig:eggbox-5-eta-heat} shows the
equilibrium density $\eta(\log T)$ after the ladder has achieved uniform
acceptance ratios.

\begin{figure}
    \begin{center}
        \includegraphics[width=\maxwidth{\columnwidth}]{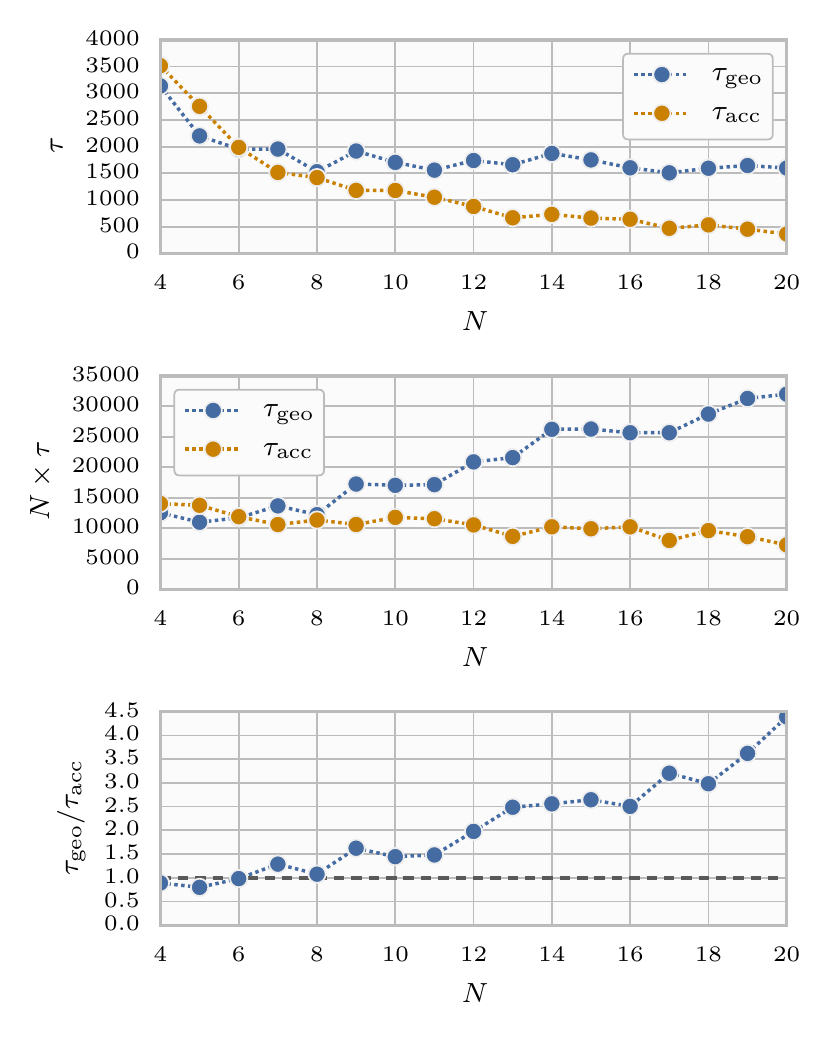}
        \caption[Autocorrelation times: eggbox]
        {\label{fig:eggbox-5-act}
            {\bf Top:} the \acp{ACT} of the cold chain ($T = 1$) of a sampler
            exploring the egg-box likelihood \eqref{eq:eggbox-5-likelihood} with
            ladders of different sizes $N$, for both geometric temperature
            ladders and ladders dynamically adapted for uniform acceptance
            ratios.
            {\bf Middle:} the total CPU time, $N \times \tau$, for the runs.
            {\bf Bottom:} the relative improvement in \ac{ACT} conferred by
            dynamically adapting for uniform acceptance ratios over a geometric
            ladder.
        }
    \end{center}
\end{figure}

\autoref{fig:eggbox-5-act} shows the \acp{ACT} of the cold chain ($T = 1$) under
uniform-$A$ and geometric ladders for the 5-dimensional egg-box problem as a
function of the number of temperatures available.  In this case, adding more
temperatures to a geometric ladder does not reduce the measured \ac{ACT} of the
sampler for $N \ge 7$, since they are added to the high-$T$ end of the ladder,
above $\Tprior$, and the ratios between lower temperatures do not change.
\autoref{fig:eggbox-5-evolution} shows that from the initial geometric ladder
only around 6 chains are within the range of temperatures spanned by the
equilibrium ladder; the remaining 8 (excluding $T_1 = 1$ and $T_N = \infty$) are
all above $\Tprior$ and effectively sample from the prior.  In this case,
therefore, the geometric spacing that would give uniform acceptance ratios of
$25\%$ for an ideal Gaussian in fact spaces temperatures too widely for $\gtrsim
6$ chains.

Meanwhile, adding more chains to a dynamically adapted ladder clearly reduces
the \ac{ACT} of the sampler in this regime.  Moreover, the \ac{ACT} of a sampler
using a uniform-$A$ ladder is always lower than that of a sampler using the
geometric ladder of the same $N$.  In the egg-box example, which requires a
relatively close spacing of temperatures, the improvement is dramatic when many
chains are used: $\actgeo > 2\actacc$ for $N \ge 12$.

The failure of the geometric ladders used in this example for $N \ge 7$ lies in
the poor $\Tmax$ chosen by assuming that the distribution behaves like an ideal
Gaussian.  A geometric spacing is in fact appropriate for a large portion of the
temperature range, but its efficacy relies on the ladder terminating at the
correct $\Tprior$.

When $N < 7$, the geometric and uniform-$A$ ladders show similar \acp{ACT}, and
the geometric ladders in fact do slightly {\em better}.  While unexpected, this
is a consequence of the behaviour of the affine invariant ensemble sampler used
in \emcee{} \citep{emcee, Goodman2010} as applied to the egg-box likelihood
\eqref{eq:eggbox-5-likelihood}.  When such a sampler is applied to a target
distribution for which the number of modes $n_m$ is greater than the number of
walkers $n_w$ used by the sampler, it behaves as though it is sampling from the
prior (albeit inefficiently).  There is therefore little benefit in having a
chain sampling as high as $\Tprior$, and so it is better -- in terms of the
\ac{ACT} -- to assign more chains to lower temperatures in order to increase
their acceptance ratios.  In our case, the egg-box likelihood has 243 modes in
5 dimensions, while the sampler uses only 100 walkers, so these walkers tend to
become isolated from one another.  Since the sampler relies on clustering of
walkers on an individual mode to inform jump proposals within that mode, jumps
are instead proposed {\em between} modes when there are on average fewer than
one walker per mode.

We anticipate that running the same tests on a traditional single-walker MCMC
sampler, or reducing the number of modes of the likelihood distribution so that
$n_w \gg n_m$, will dramatically increase $\actgeo / \actacc$ in the low
temperature regime.  We should expect that $\actgeo \gg \actacc$ when $\Tmax(N)
\ll \Tprior$ for the geometric ladder and that $\actgeo \approx \actacc$ when
$\Tmax(N) \approx \Tprior$.

\autoref{fig:eggbox-5-act} therefore illustrates a very specific case for $N < 7$
that does not reflect the importance of choosing $\Tmax \approx \Tprior$.
Nonetheless, the \acp{ACT} of the two temperature allocation strategies --
geometric and uniform-$A$ -- are still fairly similar for $N \le 7$ and there is
a distinct improvement for $N > 7$.

\section{Gravitational wave signals from compact binary coalescences}
\label{sec:gw}

We now present an example application of the dynamics of \autoref{sec:adaptive}
to a challenging and computationally expensive astrophysical inference problem.

The first direct detections of \acfp{GW} by ground-based \ac{GW} detectors
comprising the advanced LIGO and Virgo observatories \citep{aLIGO, aVirgo} are
anticipated within this decade.  Among the most promising candidate sources for
these detectors are extragalactic \acp{CBC} \citep{Rates}: the energetic
inspiral and merger of pairs of neutron stars and/or black holes, emitting a
well-modelled \ac{GW} signal.

For \ac{GW} astrophysicists, recovery of the source parameters of such a signal
from the \ac{GW} signature as observed by the detectors is a significant
challenge in the application of Bayesian statistics
\citep[e.g.,][]{vanderSluys2008, vanderSluys2008a, Raymond2010, Rodriguez2014,
S6PE, Vitale2014, Singer2014, Veitch2015}.  Parallel-tempered \ac{MCMC} is one
method used within the \ac{LSC} to perform this inference, so parameter
estimation for \ac{CBC} detections presents an ideal test for the scheme
outlined in \autoref{sec:adaptive}.

\subsection{Parameter estimation for gravitational waves}

We aim to recover the Bayesian posterior probability density $\pi(\vectheta)$
for the parameters $\vectheta$ of a merging binary system from which a \ac{GW}
signal has been detected, given some detector data $s$ and priors
$p(\vectheta)$.

The likelihood function $L(\vectheta;s)$ for this problem is a function of
\begin{enumerate}
    \item the detector output $s$, and
    \item a model waveform $h(\vectheta)$ representing a putative \ac{GW}
        signal,
\end{enumerate}
such that $s = h(\vectheta) + n$, where $n$ is the noise in the detector and
$s$, $h$, and $n$ are time series with Fourier transforms $\tilde{s}$,
$\tilde{h}$, and $\tilde{n}$ respectively.

Given a noise model for the detector, $\tilde{n}$ is then a random variable with
a known distribution and $L(\vectheta;s)$ can be defined in terms of $\tilde{n}
= \tilde{s} - \tilde{h}(\vectheta)$.  Under a simple stationary Gaussian noise
model this likelihood can be expressed as
\begin{equation}
    \label{eq:gw-likelihood}
    \logL(\vectheta;s) = -\frac{1}{2}\int_{-\infty}^\infty \frac{|\tilde{s}(f) - \tilde{h}(f;\vectheta)|^2}{S_n(f)}
        \dif{f}
        + C
    \text{,}
\end{equation}
where $C$ is a normalising constant and $S_n$ is the two-sided noise \ac{PSD}.

The approximate \ac{CBC} waveform $h$ that determines this likelihood has
between 9 and 15 physical parameters in $\vectheta$.  These parameters can be
partitioned into two disjoint groups; firstly, the intrinsic parameters are
those that determine the dynamics of the source itself:
\begin{itemize}
    \item $q$ and $\Mc$: the mass ratio and chirp mass of the binary,
        system\footnote{
            This is an alternative parameterisation of the mass configuration of
            a binary system with component masses $m_1$ and $m_2$, such that $q
            \equiv m_2/m_1$ and $\Mc \equiv (m_1 m_2)^{3/5}/(m_1 + m_2)^{1/5}$.
        },
    \item $a_1$ and $a_2$: the spin magnitudes of the binary components, and
    \item $t_1$, $t_2$, $\phi_\mathrm{JL}$, and $\phi_{12}$: the four angles
        describing the spin orientations of the binary components.
\end{itemize}

Secondly, the extrinsic parameters are those that determine only the waveform
observed at the detector:
\begin{itemize}
    \item $d_L$: the luminosity distance between the binary system and the
        detector,
    \item $\alpha$ and $\delta$: the right ascension and declination of the
        event,
    \item $\psi$ and $\theta_\mathrm{JN}$: the two angles describing the orbital
        orientation of the binary system, and
    \item $t_\mathrm{c}$ and $\phi_\mathrm{c}$: a reference time and phase for
        the waveform.
\end{itemize}

A binary system with arbitrary spinning component masses is therefore described
by 15 parameters.  A binary system whose spins are aligned with its orbital axis
is modelled by 11 parameters, neglecting the spin orientation angles, and a
non-spinning system is modelled by only 9 parameters, with the spin magnitudes
also omitted.

This likelihood function and its parameter space generate a highly structured
posterior distribution with many modes and degeneracies that is difficult for
conventional \ac{MCMC} samplers to explore.

\subsection{Dynamic parallel tempering}
\label{sec:gw-application}

We implemented the scheme of \autoref{sec:adaptive} in the \ac{MCMC} sampler
used by \lalinf{}: the software suite used by the \ac{LSC} for \ac{GW} parameter
estimation \citep{Veitch2015}.  Unlike \emcee{}, the \lalinf{} sampler uses only
one walker, with jump proposals that are tuned to the structure of the posterior
distribution generated by a \ac{CBC} signal.  \citet{Veitch2015} present a
comparison of \lalinf{}'s standard parallel tempered \ac{MCMC} implementation
with other sampling techniques, such as nested sampling \citep{Skilling2006,
Veitch2010}.  They find \ac{MCMC} to be at least competitive with the other
methods in both CPU time and wall time, and in some cases better.

To compare our implementation of \autoref{sec:adaptive} under \lalinf{} with the
default geometric temperature ladder, we tested both schemes on a number of
synthetic \ac{GW} events simulating the signals received from two different
compact binary sources.

We conduct our tests against two non-spinning prototype \ac{GW} sources: a
\ac{BNS} system and a \ac{BBH} system, detailed in \autoref{tab:gw-injections}.
For each of these prototypes, we simulate coherent detections by a network of
\ac{GW} detectors, for a range of network \acp{SNR}, by injecting the computed
\ac{GW} signal into mock Gaussian noise generated from the noise \acp{PSD} of
each detector.  We simulate a network comprising the Advanced LIGO detectors in
Hanford, Washington and Livingston, Louisiana and the Advanced Virgo detector in
Cascina, Italy, using noise \acp{PSD} that approximate the detectors' design
sensitivities \citep{aLIGOPsd, aVirgoPSD}.

The \ac{SNR} $\rho$ of a \ac{GW} detection is a proxy for the maximum log
likelihood, such that $\Delta \logL$ scales as $\rho^2/2$, where $\Delta \logL$
is the difference between the maximum log likelihood under the signal model and
the likelihood of the noise-only ($h = 0$) model.  The \ac{SNR} therefore
indicates how sharply peaked the posterior distribution will be.  Since the
\ac{SNR} of a detection can be estimated by the detection pipeline, it can also
be used to decide the $\Tmax$ used in constructing a geometric temperature
ladder for that run, against which we compare a uniform-$A$ ladder.

We attempt to recover the parameters of the injected events with the likelihood
function \eqref{eq:gw-likelihood}, using two families of frequency-domain
waveform approximants:
\begin{enumerate}
    \item \tft{}, which describes with 9 to 11 free parameters the
        post-Newtonian inspiral of two masses, optionally with spins aligned
        with the orbital axis \citep{Buonanno2009}, and
    \item \imrpp{}, which describes the full inspiral-merger-ringdown sequence
        of a \ac{CBC}, allowing for arbitrary precessing spins and having 15
        free parameters (in the \lalinf{} implementation) \citep{Hannam2014}.
\end{enumerate}
When recovering with \tft{}, we allow for aligned spins in the system, while for
both approximants, we analytically marginalise the reference phase $\phi_c$ out
of the likelihood.  For our runs, therefore, the \tft{} approximant generates a
10-dimensional parameter space, while \imrpp{} generates a 14-dimensional
parameter space.

\begin{figure*}
    \begin{center}
        \includegraphics[width=\maxwidth{\textwidth}]{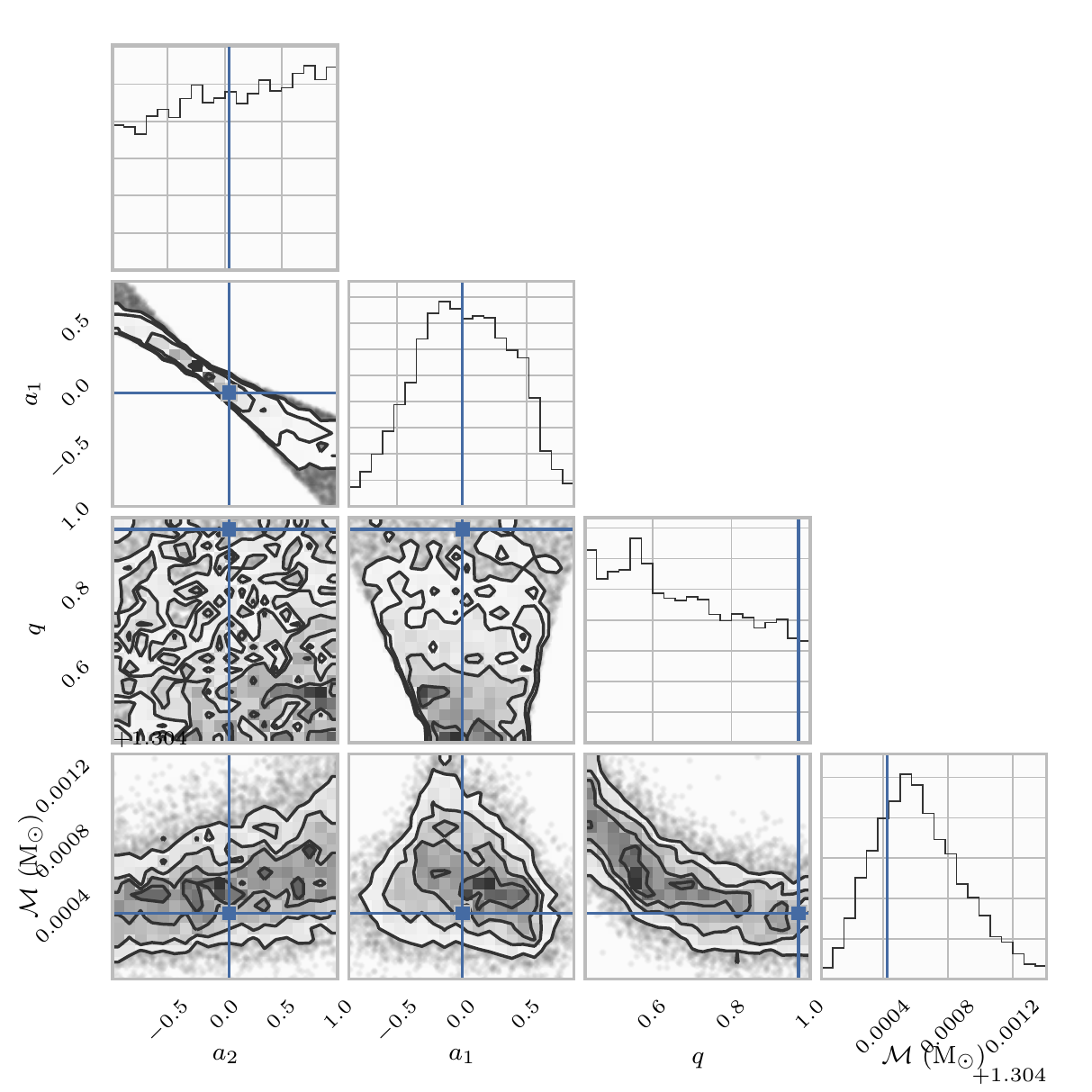}
        \caption[\lalinf{} corner plot]
        {\label{fig:lal-corner-intrinsic}
            The one- and two-dimensional marginal distributions of the intrinsic
            parameters of a \ac{BNS} event with \ac{SNR} 25.  Note, in
            particular, the very accurate measurement of the chirp mass $\Mc$
            (the plotted range is only $\sim 0.1\%$ of the true value).}
    \end{center}
\end{figure*}

\begin{figure*}
    \begin{center}
        \includegraphics[width=\maxwidth{\textwidth}]{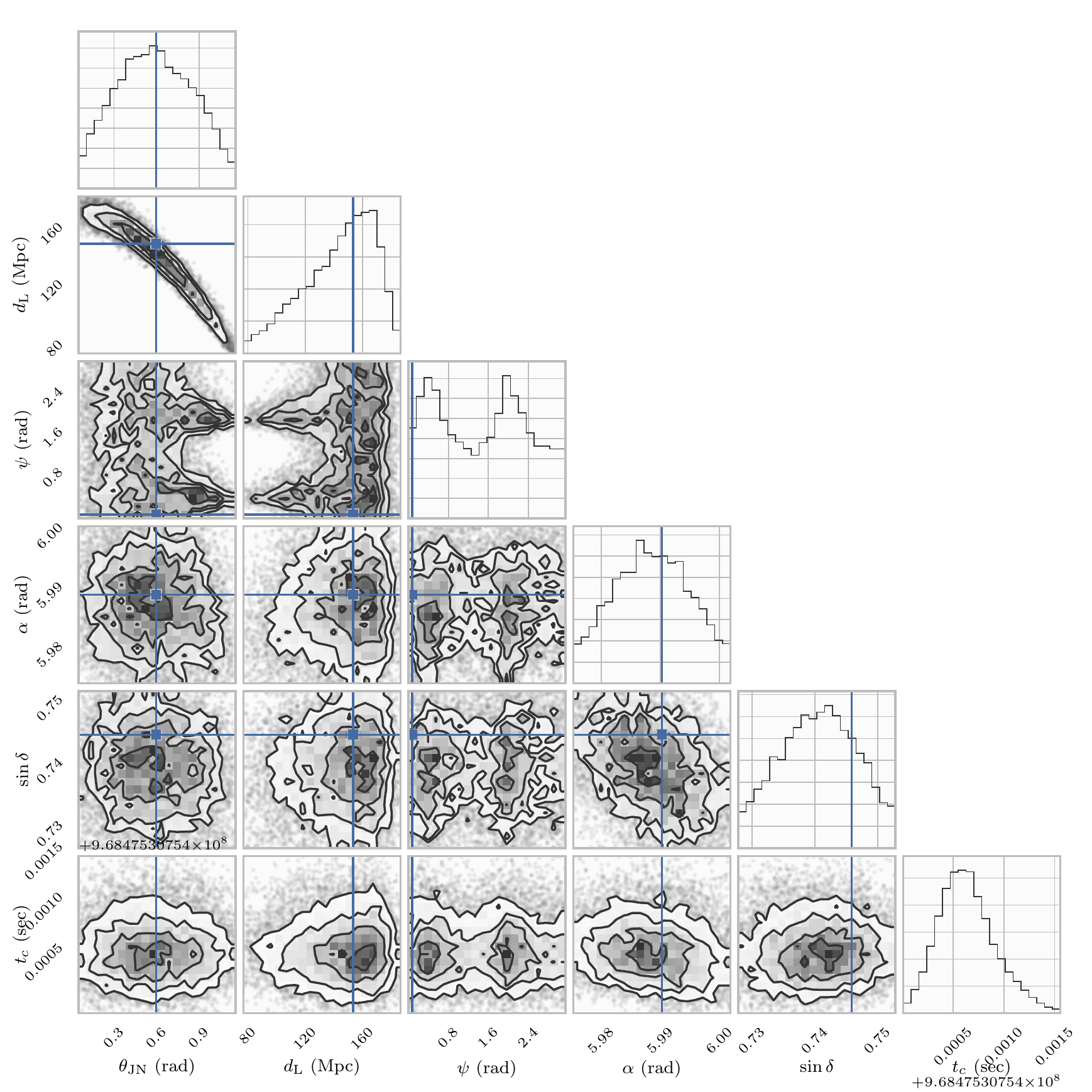}
        \caption[\lalinf{} corner plot]
        {\label{fig:lal-corner-extrinsic}
            The one- and two-dimensional marginal distributions of the extrinsic
            parameters recovered with \tft{} from a \ac{BNS} event with \ac{SNR}
            25.  Note the multiple modes for the polarization angle $\psi$ and
            the strong correlation between distance $d_\mathrm{L}$ and
            inclination $\theta_\mathrm{JN}$.
        }
    \end{center}
\end{figure*}

The posterior distribution for one of these problems, a \ac{BNS} binary
recovered with \tft{} at an \ac{SNR} of 25, is illustrated in
\autoref{fig:lal-corner-intrinsic} and \autoref{fig:lal-corner-extrinsic}.
These show the one- and two-dimensional marginal distributions of the recovered
samples, partitioned into intrinsic and extrinsic parameters.  Some parameters,
such as the chirp mass $\Mc$, are very accurately measured, while others show
multiple modes (e.g., the polarisation angle $\psi$) or strong correlations
(e.g., distance $d_\mathrm{L}$ and inclination $\theta_\mathrm{JN}$).

\begin{figure}
    \begin{center}
        \includegraphics[width=\maxwidth{\columnwidth}]{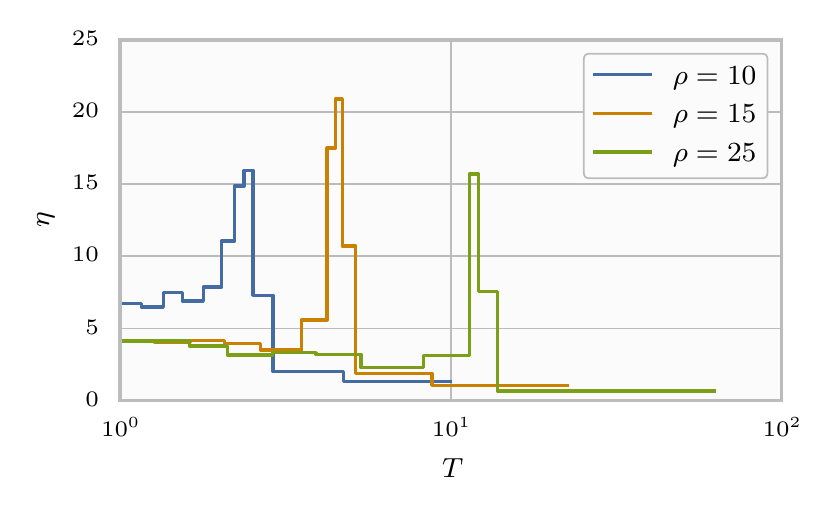}
        \caption[Chain density: BBH injection]
        {\label{fig:eta-bbh-tf2}
            The equilibrium density of chains per $\log T$, from \eqref{eq:eta},
            for the \tft{} \ac{BBH} runs described in
            \autoref{sec:gw-application} at various \acp{SNR}.
        }
    \end{center}
\end{figure}

\autoref{fig:eta-bbh-tf2} shows the effect of the \ac{SNR} on the equilibrium
(uniform-$A$) chain density $\eta$ selected by our dynamical scheme.  While the
structure of the temperature ladder is preserved, its features scale to higher
temperatures as the \ac{SNR} of the injected signal increases while the average
value of $\eta$ falls.  Since the maximum log likelihood scales with the square
of the \ac{SNR}, it follows that, under a fixed prior, $\Tprior$ should also
increase with the \ac{SNR}.

\begin{figure}
    \begin{center}
        \includegraphics[width=\maxwidth{\columnwidth}]{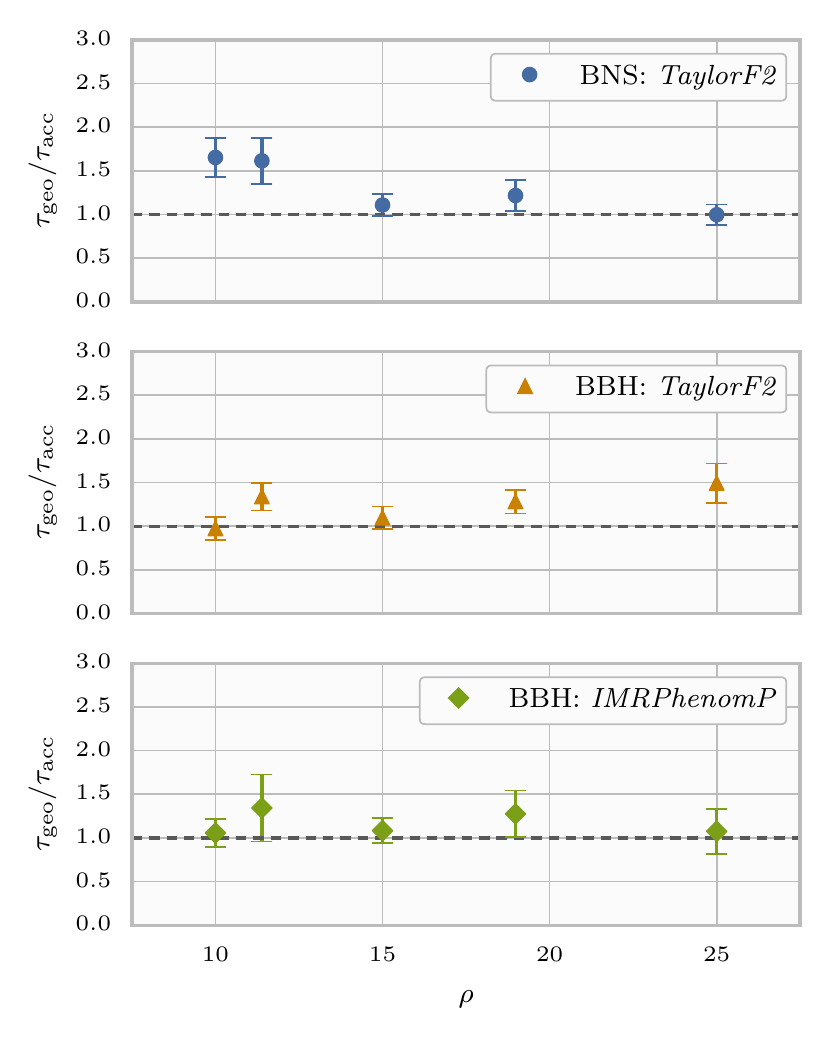}
        \caption[\lalinf{} autocorrelation times]
        {\label{fig:lal-acts}
            The fractional improvements in \ac{ACT} conferred by a uniform-$A$
            temperature ladder over a geometric ladder for the \ac{CBC}
            parameter estimation problem described in
            \autoref{sec:gw-application} at various \acp{SNR}.
        }
    \end{center}
\end{figure}

Meanwhile, \autoref{fig:lal-acts} shows the ratios of \acp{ACT} for runs using
uniform-$A$ ladders versus those using geometric ladders.  The lowest \ac{SNR}
that we simulate, 10, represents a signal that is on the threshold of
detectability, where we expect most detections to occur, while the maximum, 25,
represents a relatively loud signal (at around the $90^{\mathrm{th}}$ percentile
of detectable events).

While there is significant variation in the \ac{ACT} measurements between
\acp{SNR}, there is on average a reduction in \ac{ACT} of $26\%$ for the systems
and \acp{SNR} tested.  In general, a uniform-$A$ ladder is at least as effective
as a geometric ladder in all cases; that is, the \ac{ACT} ratio $\actgeo /
\actacc$ is never less than one (within error bars).  In some cases, this ratio
is appreciably greater than one, e.g., for low-\ac{SNR} \ac{BNS} events.

However, as we shall discuss briefly in \autoref{sec:discussion-caveats}, the
single-walker nature of the \lalinf{} sampler inhibits communication between the
extremal chains, so that the temperatures are in fact partitioned into two
independent, non-communicating groups.  The improvement we observe in
\autoref{fig:lal-acts} therefore arises in fact from more efficient allocation
of the temperatures below the critical temperature of the phase transition.
Meanwhile, those chains above the critical temperature -- which are sampling in
the regime where the noise-only model is preferred over the presence of a
\ac{GW} signal -- remain isolated.

We anticipate that, with an ensemble sampler with similar problem-specific
optimisations to those used by \lalinf{}, the communication barrier at the phase
transition could be removed and we may observe greater improvements in \ac{ACT}
and sampling efficiency.

\begin{table*}
    \begin{center}
        \caption[Compact binary sources]
        {\label{tab:gw-injections}
            The \ac{CBC} event prototypes used to test the adaptation scheme of
            \autoref{sec:adaptive}.  All prototypes are simulated at distances
            that yield 5 different \acp{SNR}: 10, 11, 15, 19, and 25.
        }
        \begin{tabular}{llccl}
            \toprule
            Source   & Injection waveform & $q$              & $\Mc$ ($\Msun$) & Recovery waveforms\\
            \midrule
            BNS      & \sttf              & $0.970$          & $1.30$          & \tft \\
            BBH      & \imrpp             & $0.996$          & $4.82$          & \tft, \imrpp \\
            \bottomrule
        \end{tabular}
    \end{center}
\end{table*}

\section{Discussion}
\label{sec:discussion}

The temperature selection scheme set out in \autoref{sec:adaptive} solves two
problems in the application of parallel tempering:
\begin{enumerate}
    \item It identifies $\Tmax = \infty$ as a suitable temperature for the hot
        chain -- such that it will sample from the prior -- that is independent
        of the target distribution.
    \item It allocates a fixed number of intermediate temperatures to ensure good
        communication between fixed extremal temperatures $\Tmin$ and $\Tmax$,
        and therefore efficient sampling of the target distribution -- i.e.,
        with few iterations between independent samples.
\end{enumerate}

The intermediate temperatures are allocated so that acceptance ratios for swaps
proposed between neighbouring pairs of chains are uniform across the temperature
ladder.  The dynamical algorithm that implements this scheme requires only two
parameters: $\nu$ and $t_0$.  These parameters, discussed in
\autoref{sec:adaptive-params}, describe only the initial dynamics of the
temperatures, setting the time-scale for temperature adjustments, and do not
determine the equilibrium uniform-$A$ ladder.

While a temperature configuration that is selected for uniform acceptance ratios
between all chain pairs is not necessarily {\em optimal} in the \ac{ACT} of the
sampler, we have demonstrated that it is generally better than a conventional
geometric temperature configuration and provides more consistent behaviour
across different likelihood distributions and numbers of chains.  Importantly,
the dynamics that achieve such a temperature ladder are simple and easily
implemented, requiring very little tuning or intervention.

The factor by which the \ac{ACT} is reduced by the uniform-$A$ scheme depends
strongly on the likelihood distribution that is explored and on the specific
geometric ladder against which the uniform-$A$ scheme is being compared.  For a
geometric ladder, one must make an ad hoc choice of the maximum temperature
$\Tmax$; this is difficult and a poor guess can yield a very sub-optimal ladder.
In particular, if $\Tmax$ is not high enough that the sampler can efficiently
migrate between modes, then the \ac{ACT} will be significantly higher than it
needs to be.  On the other hand, if $\Tmax$ is too high, then many of the chains
will effectively sample from the prior, and CPU time will be wasted in
sampling from redundant tempered likelihood distributions.

The uniform-$A$ temperature dynamics guarantee that, for a given number of
chains $N$, no such wastage of CPU time occurs and that there will always be
precisely one chain sampling at $\Tmax = \infty$ (i.e., sampling from the
prior).  Tests of the dynamics generally demonstrate lower \acp{ACT} when
compared with geometric temperature ladders of the same number of chains, $N$.

In \autoref{sec:tests-2banana-improvement} we demonstrated that, even with a
judicious choice of $\Tmax$ that is close to $\Tprior$, a traditional geometric
ladder is outperformed by a ladder chosen for uniform acceptance ratios (with
$\Tmax = \infty$).  \autoref{fig:2banana-act} illustrates that, when $\Tprior$
is known, a uniform-$A$ ladder confers the greatest reduction in \ac{ACT} when
$N$ is small.  In this case, the temperature ratio $\gamma$ of the geometric
ladder is large enough that phase transitions in the distribution of $\logL$
cause a bottleneck in the communication between hot and cold chains around a
critical temperature, where $A \ll 1$.  The uniform acceptance scheme allocates
more chains over these temperature regimes in an effort to optimise the
communication.

For larger $N$, $\actgeo / \actacc \approx 1$, suggesting that -- as long as
there are no pairs of chains with prohibitively low swap acceptance ratios -- a
geometric spacing is adequate if $\Tmax$ is chosen appropriately.

It is unclear how to determine the threshold $A$ below which communication is
impeded, but it is likely related to the time-scale of the intra-chain motion of
the sampler.  If intra-chain jumps are accepted seldom with respect to the rate
of inter-chain swaps, then increasing the inter-chain swap acceptance ratio is
unlikely to make the sampler any more efficient.

In general, while $\actgeo > \actacc$ for all $N$, the improvement fraction
$\actgeo/\actacc$ will asymptote to 1 as $N \to \infty$.  The rate of decay will
depend strongly on the target distribution.  A system with a wide distribution
of $\logL$ (e.g., with many dimensions) or with sharp phase transitions at
certain temperatures (e.g., with many modes of various shapes and weights) will
see the most benefit from having many chains, while a better-behaved
distribution without such features can be efficiently sampled with fewer.

Meanwhile, from our tests on the 5-dimensional egg-box distribution discussed in
\autoref{sec:tests-eggbox}, we can see the consequences of a poor choice of
$\Tmax$.  While the egg-box distribution does not have as strong a phase
transition as the double Rosenbrock function of \autoref{sec:tests-2banana}, our
ignorance of $\Tprior$ means that a geometric ladder (which in this case is
constructed from a fixed temperature ratio $\gamma$) is mostly worse than a
uniform-$A$ ladder.  \autoref{fig:eggbox-5-act} demonstrates this, specifically
when $N$ is large enough that for a given $\gamma$ the geometric ladder places
many temperatures redundantly above $\Tprior$.  In this case, we see a dramatic
improvement in \ac{ACT} $\tau$ from using a uniform-$A$ ladder when compared
with a geometric ladder of the same number of chains $N$; indeed, the ratio
$\actgeo/\actacc$ becomes as large as $\sim 4$ for the values of $N$ tested.
Since $\actgeo$ is independent of $N$ when $N \gtrsim 7$, we should expect that
this ratio will saturate as $N \to \infty$, where $\actacc$ reaches a minimum.
Moreover, the CPU time, $N \times \tau$ of the uniform-$A$ runs continues to
decrease with $N$ in the explored range, even as the CPU time of the geometric
runs rises.

On the other hand, when $N$ is too small for a geometric ladder to reach the
prior (i.e., $T_N \ll \Tprior$), we notice that in fact $\actgeo < \actacc$.  As
discussed in \autoref{sec:tests-eggbox}, this somewhat surprising result arises
from a limitation of the ensemble sampler that was used to sample the
distribution.  We anticipate that if the number of walkers were increased to
many times the number of modes -- which is required for efficient sampling --
the geometric ladder will fail dramatically in this regime of $N$, giving
$\actgeo \gg \actacc$.

Finally, our astrophysical application in \autoref{sec:gw} illustrates the use
of the scheme of \autoref{sec:adaptive} in a single-walker setting.  In this
case, a uniform-$A$ ladder improves on the default geometric ladder used
by \lalinf{}, giving on average a $26\%$ reduction in \ac{ACT}.  These tests do,
however, reveal an instability in the uniform-$A$ dynamics under a single-walker
sampler, which we describe below.

\subsection{Single-walker implementation}
\label{sec:discussion-caveats}

The dynamical algorithm set out in \autoref{sec:adaptive} can be implemented in
a traditional MCMC sampler that uses only a single sample per chain, with
reductions observed in the \ac{ACT} of the sampler relative to a geometric
temperature ladder, as shown in \autoref{sec:gw-application}.  However, in this
case, equal (and large) acceptance rates between all chains do not guarantee
good communication of sample positions between extremal temperatures.

A complex, multidimensional posterior such as that described in
\autoref{sec:gw-application} may exhibit a phase transition.  At $T=1$, the
posterior is dominated by the likelihood peaked around the true parameter
values; at $T=\infty$, it is dominated by the much larger prior volume far away
from the parameter values, corresponding to a weak or absent signal; and at the
phase transition temperature, the two contribute comparably to the posterior,
which is distinguished by two distinct likelihood peaks: a high-likelihood peak
near the signal parameters, and a low-likelihood peak in the region of
significant prior support.  In effect, this becomes a reversible-jump \ac{MCMC}
problem with two distinct modes: the high-likelihood signal model, and the
low-likelihood noise-only model.

The dynamical algorithm with a single walker has a tendency to select very small
temperature gaps around phase transitions.  From \eqref{eq:swap}, when
$\Delta\beta \to 0$, $A \to 1$ regardless of the likelihoods of the chains.
However, the likelihood distributions near the phase transition temperature will
be distinct enough that there is no intra-chain migration of samples between
them.  Consequently, despite efficient communication between chains, the higher
temperatures do not help low-temperature walkers to efficiently jump between the
two modes, since the high- and low-likelihood modes do not mix well at any
temperature.

\begin{figure*}
    \begin{center}
        \includegraphics[width=\maxwidth{\textwidth}]{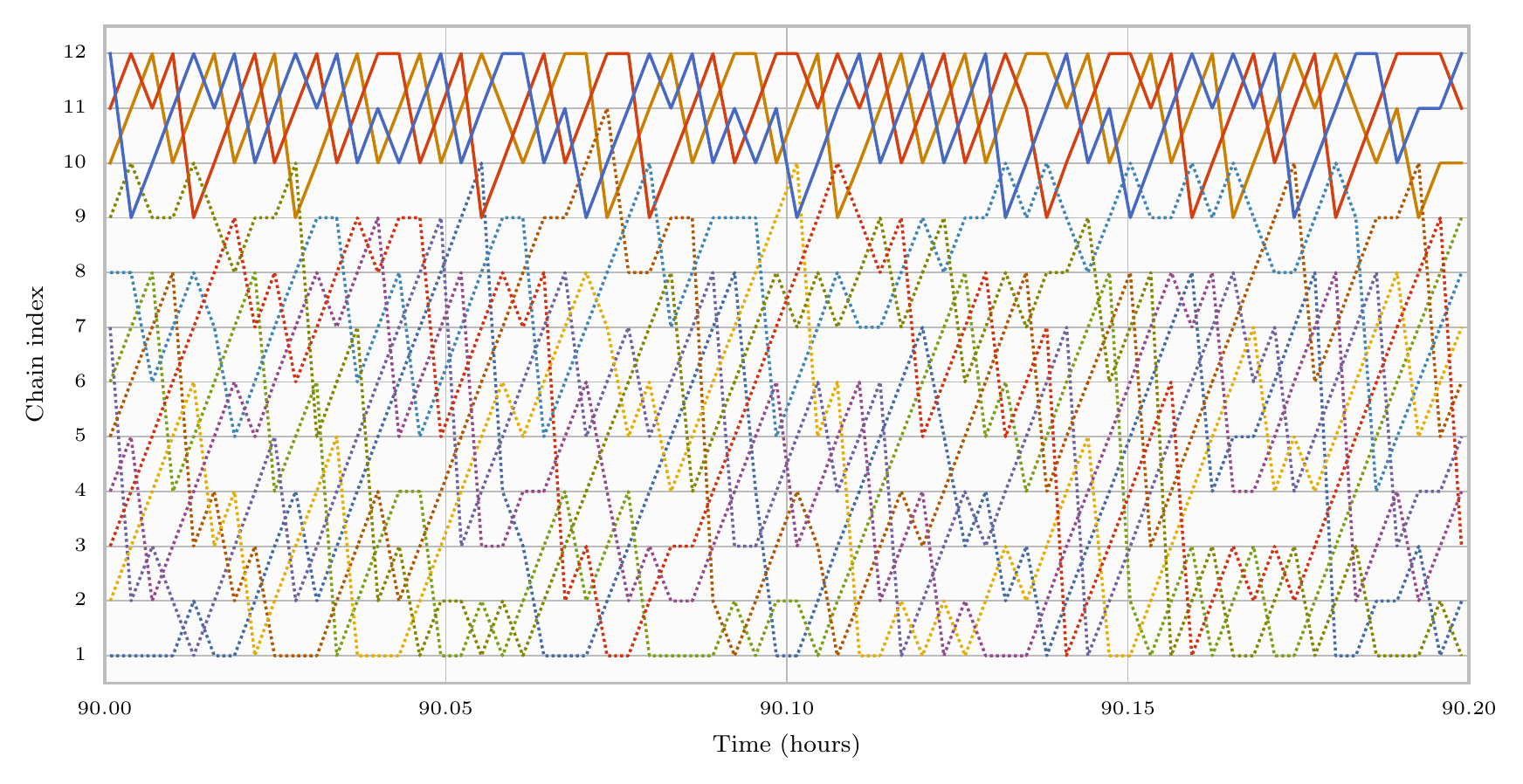}
        \caption[\lalinf{} sample paths]
        {\label{fig:paths}
            The paths traced out between temperatures by the 12 samples in a
            single-walker run with \lalinf{} on a \ac{BNS} signal of \ac{SNR}
            15.  Samples are identified by their colour.  While swap proposals
            between chains 9 and 10 are freqently accepted, there is no
            migration of samples starting above chain 10 (solid lines) to chains
            below 9, and vice versa (dotted lines).
        }
    \end{center}
\end{figure*}

\autoref{fig:paths} demonstrates how large acceptance ratios do not guarantee
the transmission of sample positions between low and high temperatures when
there is only one sample per temperature.  In this case, 9 samples (whose paths
are shown as dotted lines) occupy the high-$\logL$ part of the parameter space
representing the signal model, while the remaining 3 (solid lines) occupy the
low-$\logL$ part, representing the noise-only model.  This barrier occurs at the
temperature at which the evidence for the noise-only model is equal to that of
the signal model.

We did not encounter this issue with an ensemble sampler, which is less
susceptible to this instability because it allows the ensemble at each
temperature to occupy both models simultaneously.

\subsection{Evidence calculations}
\label{sec:discussion-evidence}

The current paper focuses mainly on the efficiency of a parallel tempered
\ac{MCMC} sampler in producing independent samples from its target distribution.
Another important task in Bayesian statistical inference is to compute the
evidence integral of the posterior distribution.  At a given temperature, this
is given by
\begin{equation}\label{eq:evidence}
    Z(\beta) \equiv \int L(\vectheta)^\beta p(\vectheta) \dif{\vectheta}
    \text{,}
\end{equation}
where $\beta \equiv 1/T$ is the inverse temperature.

Since we are interested in the untempered posterior, we wish to calculate
$Z(1)$.  From \eqref{eq:evidence}, we can use thermodynamic integration
\citep{Goggans2004, Lartillot2006} to express the log evidence (relative to the
prior) in terms of the mean $\logL$, such that
\begin{equation}\label{eq:thermodynamic-evidence}
    \Delta\log Z \equiv \log Z(1) - \log Z(0) = \int_0^1 \E[\beta]{\logL} \dif{\beta}
    \text{,}
\end{equation}
to which the logarithm of the integral of the prior, $\log Z(0)$, can be added
to give the absolute evidence $\log Z(1)$.

The log evidence can therefore be computed by a sampler through numerical
integration of the mean $\logL$ values collected over all of the chains.  In the
same way that inter-chain communication is hindered by phase transitions in the
system, numerical estimation of this integral is susceptible to sharp changes in
$\logL$ with the temperature $T$.  Such phase transitions are marked by a
diverging specific heat $C_V$ since, from \eqref{eq:specific-heat}, $C_V$ is the
derivative of $\logL$ with respect to $T$.

Since allocating temperatures for uniform acceptance ratios yields a logarithmic
chain density $\eta$ that appears to scale with $\sqrt{C_V}$, such a temperature
ladder will naturally increase the accuracy of numerical estimates of
\eqref{eq:thermodynamic-evidence} with respect to one that does not increase
$\eta$ around phase transitions.

We can test the degree of improvement conferred by a uniform-$A$ ladder by
returning to the truncated Gaussian discussed in \autoref{sec:tests-gauss}.
Normalising \eqref{eq:gauss-posterior} so that $\max \logL = 0$,
the log evidence is
\begin{equation}
    \begin{split}
        \Delta\log Z &= \left( \frac{\sqrt{2}}{R} \erf\left(\frac{R}{\sqrt{2}}\right) \right)^n \Gamma\left(1 + \frac{n}{2}\right) \\
                     &\approx -55.1
        \text{,}
    \end{split}
\end{equation}
with $R = 30$ and $n = 25$.

\begin{figure}
    \begin{center}
        \includegraphics[width=\maxwidth{\columnwidth}]{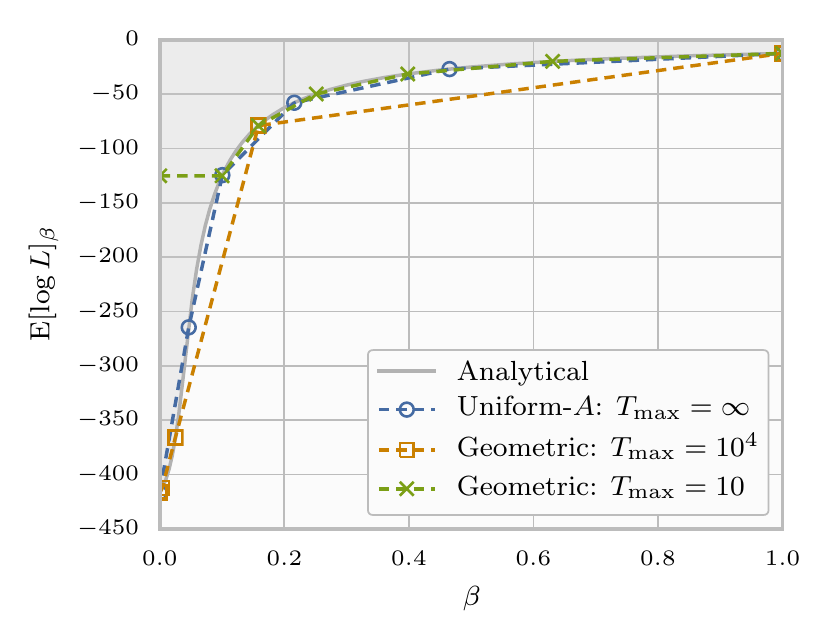}
        \caption[Gaussian evidence approximation]
        {\label{fig:gauss-evidence}
            An illustration of the thermodynamic quadrature estimates of the log
            evidence of the truncated Gaussian discussed in
            \autoref{sec:tests-gauss}.  The shaded area shows the analytical
            mean $\logL$ as a function of $\beta$, while the dashed lines
            illustrate numerical approximations from the values of $\logL$
            collected by samplers with different ladders, each of 6
            temperatures.  The resulting evidence estimates are reported in
            \autoref{tab:evidence}.  Note the denser spacing of temperatures in
            the high-curvature region for the uniform-$A$ ladder, and the errors
            incurred in extrapolating from $\Tmax$ to $T=\infty$ ($\beta=0$) for
            the geometric ladders.
        }
    \end{center}
\end{figure}

\autoref{fig:gauss-evidence} illustrates the numerical estimates of $\Delta\log
Z$ from a uniform-$A$ ladder (with $\Tmax = \infty$) and from geometric ladders
with $\Tmax = 10$ and $\Tmax = 10^4$.  The evidence quadratures for the
geometric ladders are augmented with a copy of $\E[\Tmax]{\logL}$ placed at $T =
\infty$ as a crude measure to cover the integration domain.

The evidence estimates recovered from these samplers are reported in
\autoref{tab:evidence} for 6 chains and 10, from which it is clear that
selecting temperatures for uniform acceptance ratios can greatly increase the
accuracy of the evidence estimate, bypassing the need to select a good initial
temperature ladder.  Note that the under- and over-estimates of $\Delta\log Z$
from the geometric ladders in this case are a consequence of poor choices of
$\Tmax$ rather than of sharp changes in $\E{\logL}$.  While these comparisons
are reasonable -- since for a geometric ladder it is very difficult to pick an
appropriate $\Tmax$ in advance -- we expect the presence of phase transitions to
increase this disparity, and with it the advantages of adapting the ladder
dynamically for uniform acceptance ratios.

% Values from code:
%
%               N=6                     N=10
% Adaptive:     -57.96824663984206      -55.86100491559934
% Over:         -78.00673046585639      -61.75372811503053
% Under:        -42.34219597739395      -41.62473456204515
%
% Analytical:   -55.10551862830987
\begin{table}
    \begin{center}
        \caption[Evidence estimates]
        {\label{tab:evidence}
            The evidence values of the truncated Gaussian of
            \autoref{sec:tests-gauss}, estimated from a samplers of 6 and 10
            temperatures allocated in three different ways, as compared to the
            analytical result.
        }
        \begin{tabular}{lcc}
            \toprule
                                                       & \multicolumn{2}{c}{$\Delta\log Z$} \\
            \cmidrule(l){2-3}
            Temperature ladder                         & $N = 6$ & $N = 10$ \\
            \midrule
            Uniform-$A$: $\Tmax = \infty$              & -58.0   & -55.9 \\
            Geometric: $\Tmax = 10^4$                  & -78.0   & -61.8 \\
            Geometric: $\Tmax = 10$                    & -42.3   & -41.6 \\
            \midrule
            Analytical result                          & \multicolumn{2}{c}{-55.1} \\
            \bottomrule
        \end{tabular}
    \end{center}
\end{table}

\subsection{Other measures of optimality}
\label{sec:discussion-optimality}

We have investigated the performance of a temperature ladder adapted for uniform
acceptance ratios in reducing the \ac{ACT} of a parallel tempered \ac{MCMC}
sampler.  The \ac{KL} divergence discussed in \autoref{sec:pt-ideal-gaussian}
provides an alternative measure of the distance between two temperatures.  In
\autoref{sec:tests-gauss} we showed that uniform \ac{KL} divergence in a
temperature ladder does not correspond to uniform acceptance ratios beyond the
special case of the ideal, unbounded Gaussian distribution described in
\autoref{sec:pt-ideal-gaussian}.

When applied to the truncated Gaussian discussed in \autoref{sec:tests-gauss},
for which $\KL(\pi_{T_i}\|\pi_{T_{i+1}})$ is analytically available, the $\KL$
and $A$ measured between chains drop off at different rates as $T$ approaches
$\Tprior$ (see \autoref{fig:kl-gauss-contour}).  Indeed, it is possible in
principle to estimate the temperature-dependent normalising constants required
to adapt on the \ac{KL} divergence \citep{Geyer1994, Cameron2014}.  It may be
interesting to investigate such a scheme for resilience to the single-walker
instability mentioned in \autoref{sec:discussion-caveats}.

Meanwhile, \citet{Katzgraber2006} propose an optimisation scheme in which
temperatures are chosen to minimise the round-trip time of a sample from $\Tmin$
to $\Tmax$, which they suggest will improve sampling performance on systems with
strong phase transitions.  Their algorithm is tested on simulations of the
two-dimensional Ising model, and is shown to select a different temperature
configuration than the uniform-$A$ scheme that has been discussed so far.

However, the \ac{ACT} of the sampler -- what we are ultimately concerned with in
efficient Bayesian inference -- is not discussed, so it is unclear whether this
strategy is better than selecting temperatures for uniform acceptance ratios.
Their feedback optimisation method in fact prefers a higher density of chains
per $T$ across phase transitions of the system than the uniform-$A$ scheme.  We
have shown, however, that the \ac{ACT} yielded by a particular ladder is not
critically sensitive to under-densities over phase transitions so long as the
acceptance ratio is not prohibitively small in these temperature regimes (see
\autoref{sec:tests-2banana-improvement}).  Indeed, increasing the density of
chains over phase transitions too far might unnecessarily hinder inter-chain
communication at other temperatures (by reducing $A$), leading to an overall
rise in \ac{ACT}.

These reservations, together with the complicated book-keeping involved in
optimising for round-trip time, lead us to favour the dynamical method presented
in \autoref{sec:adaptive}.  By comparison, this dynamical method is simple and
guaranteed to produce an equilibrium ladder that yields efficient -- if not
perfectly optimal -- sampling from any target distribution, with the proviso of
many walkers per temperature.

\section*{Acknowledgements}

We thank Ewan Cameron, Christopher Berry, and Daniel Foreman-Mackey for their
critique and assistance.  \autoref{fig:lal-corner-intrinsic} and
\autoref{fig:lal-corner-extrinsic} were generated with \textsc{triangle.py}
\citep{triangle}.  This work was supported by the Science and Technology
Facilities Council and the Leverhulme Trust research project grant.  We are
grateful for computational resources provided by Cardiff University, and funded
by an STFC grant supporting UK Involvement in the Operation of Advanced LIGO.

A reference implementation of the algorithm, based on the ensemble sampler
\emcee{} \citep{emcee}, is available at \emceeref.  

\bibliographystyle{mnras}
\bibliography{paper}

\appendix

% vim: set ft=tex :

\section{Autocorrelation time estimation}
\label{sec:act-estimation}

The \acf{ACT} discussed in this paper refers to the {\em integrated
autocorrelation time} described by \citet{Sokal1997}.  It is estimated in the
following way.

If $x(t)$ is a time series with a normalised autocorrelation function $\rho(t)$,
such that $\rho(0) = 1$, then the integrated \ac{ACT} of $x$ is defined by
\begin{equation*}\label{eq:act-def}
    \begin{split}
        \tau &\equiv \sum_{t = -\infty}^{\infty} \rho(t) \\
             &= 1 + 2\sum_{t=1}^{\infty} \rho(t)
                \text{.}
    \end{split}
\end{equation*}

Since, when $t \gg \tau$, $\rho(t) \approx 0$, there is little contribution to
the integral at large lags, except through noise in the measured autocorrelation
function $\rho$.  We can therefore approximate the \ac{ACT} as
\begin{equation*}\label{eq:act-estimate}
    \tau \approx 1 + 2\sum_{t=1}^{M\tau}\rho(t)
    \text{.}
\end{equation*}

We estimate the \ac{ACT} over a window that is $M=5$ \acp{ACT} long, subject to
the constraint that $M\tau < N/2$, where $N$ is the number of samples in $x$.
If this constraint is violated, the result is probably not trustworthy, since
there are too few samples for a meaningful estimate.

\label{lastpage}

\end{document}